\documentclass[]{interact}
\usepackage{subfigure}
\usepackage{amsmath}
\usepackage{soul,color}
\usepackage{graphicx}
\usepackage{dcolumn}
\usepackage{bm}
\usepackage{times}
\usepackage[numbers,sort&compress,merge]{natbib}
\renewcommand\bibfont{\fontsize{10}{12}\selectfont}
\renewcommand\bibnumfmt[1]{(#1)}

\theoremstyle{plain}

\theoremstyle{definition}

\theoremstyle{remark}

\begin{document}

\title{Quantum thermometry in a squeezed thermal bath}

\author{
\name{H. Rangani Jahromi \thanks{ Email: h.ranganijahromi@jahromu.ac.ir}}
\affil{Physics Department, Faculty of Sciences, Jahrom
University, P.B. 7413188941,  Jahrom, Iran.}
}

\maketitle

\begin{abstract}
We address  the dephasing dynamics of the quantum Fisher information (QFI)  for the process of quantum thermometry  with probes coupled to 
   squeezed thermal baths via the nondemolition interaction.  We also calculate the upper bound for the parameter estimation and investigate how the optimal estimation is affected by the initial conditions and decoherence, particularly the squeezing parameters.
   Moreover,  the feasibility of the optimal measurement of the temperature is discussed in detail.
   Then, the results are generalized  for entangled probes and  the multi-qubit scenarios for probing the temperature are analysed. 
   Our results show that 
   the squeezing can  decrease 
   the number of channel uses for optimal thermometry. Comparing  different schemes for multi-qubit estimation, we find  that an increase in the number of the qubits, interacting with the channel, does not necessarily vary
   the precision of estimating the temperature.  Besides, we discuss the enhancement of 
  the quantum thermometry using the parallel strategy and starting from the W state.

\end{abstract}


\section{Introduction\label{introduction}}

Acquiring information about the world is realized by observation and measurement, and the results of which are subject to error \cite{Helstron}. The classical approach to decrease the statistical error is increase in the
the  resources for the measurement according to
the central limit theorem; but this method is not always 
desirable or efficient \cite{Israel2014}.
Quantum parameter estimation theory describes strategies allowing the
estimation precision to surpass the limit of classical approaches
\cite{GiovannettiPRL2006,Giovannetti2004,Dam2007,M.G.A. Paris,RanganiAOP,RanganiAOP2,RanganiJMO,RanganiOPTC,RanganiQIP4,RanganiAr1}. When the quantum system is sampled N times,  different
strategies \cite{Huang2018} allowing one to achieve the Heisenberg
limit, can be designed  such that the variance of the estimated parameter scales
as $ 1/N^{2} $.
Initially entangled probes, however, may in principle offer a significant enhancement in precision of parameter estimation \cite{Huang2018,Demkowicz2012,Berry2000,Zwierz2010,DemkowiczPRL2014,Qin LiuNCOM}. 
Those strategies have been  realized experimentally in
atomic spectroscopy \cite{Wineland1992,Huelga1997} in which the spin-squeezed
states have been employed for improving frequency calibration
precision \cite{Wasilewski2010,Koschorreck2010}. 
 Besides, the same quantum enhancement principle
can be utilized in optical interferometry \cite{Mitchell2004,Nagata2007} with exciting applications in the process of seeking the first direct detection of gravitational
waves \cite{LIGO}.

On the other hand, it has been also proven the entanglement is not always 
 useful for parameter estimation \cite{Hyllus2010} and  there are  highly entangled
 pure states that are not useful in the process of quantum estimation. Moreover, it has been discussed \cite{GiovannettiPRL2006} while entanglement at the initial
 stage  can be useful to enhance the precision of the estimation,
 entangled measurements are never necessary. Besides, there are some cases in which if
  the probes are initially prepared in an  unentangled state,  a better performance for the parameter estimation is attainable \cite{Boixo2008}. 
   Specifically, in Ref. \cite{Huang2018} it has been illustrated   that in the presence of
   Pauli x-y, depolarizing, and amplitude damping noise, unentangled probes perform better  in the
      high-noise regime.
Particularly,
  the view
that  entanglement is necessary for quantum-enhanced metrology
has been challenged in Refs. \cite{Tilma2010,Datta2012} by demonstrating that
the enhancement, obtained via entanglement, may be contingent on the
final measurement  and the way in which the unknown
parameter is encoded into the probe quantum state.
 In addition,  it has been
illustrated that under certain conditions the entanglement may even lead to
deleterious effects in quantum metrology \cite{Sahota2015}.
According to the above  discussion, presenting a universal prescription, applicable for all quantum systems,  about the relation between the QFI and entanglement, is not possible. These reasons  motivate us to investigate more the role of the entanglement in  different quantum scenarios for quantum metrology.

\par

Recently, the quantum parameter estimation theory have attracted increasing attention
in the field of quantum thermodynamics in which  accurate estimation and
control of the temperature are very significant \cite{Williams2011,Kliesch2014,Millen2016,Vinjanampathy2016}.
In addition to 
the emergence of primary and secondary thermometers based
on precisely machined microwave resonators \cite{Mohr2005,Weng2014}, recent
studies have been focused on measuring temperature at
even more smaller scales in which nanosize thermal baths are extremely
sensitive to disturbances induced by the probes \cite{Pasquale2016,Pasquale2017,Rangani2018,Campbell2017,Palma2017}.
 Some interesting paradigms of nanoscale thermometry are quantum harmonic oscillators \cite{Brunelli2012}
nanomechanical resonators \cite{Brunelli2011}, and atomic condensates \cite{Johnson2016,Hohmann2016}. 
On the other hand, temperature plays an important role in realizing phase-matching condition in non-linear optical materials. Besides,
thermal processes may result in large  nonlinear optical
effects originated from temperature dependence of the material  refractive index \cite{BOYD}.
Accordingly, precise determination of the temperature is of great importance in all branches of
modern science and technology \cite{P. Neumann2013,G. Kucsko2013,Toyli2013}. 
A scheme for enhancing the sensitivity of quantum thermometry is proposed in \cite{Kiilerich2018} where the sensing quantum
system used to estimate the temperature of an external bath is dynamically coupled with an external ancilla
(a meter) via a Hamiltonian term. 
Moreover, the dephasing dynamics of a single-qubit as an effective process in order to estimate the temperature of its
environment is addressed in \cite{Razavian2018}.  Here, we generalize the results by using the entangled probes in the presence of squeezed noises \cite{Klaers2017} which are   of great importance for the thermometry of the thermal bath.
In particular,  investigating the quantum metrology in the presence of these quantum noises completes other studies focusing on   exploring the characterization of complex environments described
 classically \cite{Benedetti0321142014,Benedetti24952014,Rossi0103022015,Javed172018,Kenfack11232019}.

\par
 One of the main design considerations in digital electronics  is energy dissipation.  According to  \textit{Landauer's principle} \cite{Landauer}, the
erasure (or reset) of one bit of classical information is
necessarily associated with  an energy input of at least \text{$ k_{B} T \text{ln} 2$} and 
an entropy increase of at least \text{$ k_{B} \text{ln} 2$}. On the other hand, it is
predicted that the Landauer limit will be achieved within
the next few decades \cite{Pop2010}. Therefore, improving our
understanding of energy dissipation in information processing devices are of both theoretical and  technological interest. Because of the advancing miniaturization, nonequilibrium and quantum effects must be
also taken into account \cite{Goold2015,Manzano2018}.  Moreover, it has been demonstrated \cite{Klaers2019} that memory devices embedded in a \textit{squeezed thermal reservoirs} \cite{Scully} are unbounded by the Landauer limit.
In such environments, thermal fluctuations exhibit fast
periodic amplitude modulations, which can be used
to decrease the minimum energy costs for an erasure operation below the standard Landauer limit.
 This setup can
naturally arise in digital electronic circuits operating in a
pulse-driven fashion and, in the future, might be exploited to
build more energy-efficient electronic devices. Besides, in the context of heat engines, it has been suggested that \cite{Rossnagel2014}  squeezed
thermal states may be exploited as an additional resource to
overcome the standard \textit{Carnot limit} \cite{Carnot} bounding the efficiency of heat
engines; particularly, in Ref. \cite{Klaers2017}  a nanobeam heat engine coupled to squeezed thermal noise  has been realized, whose efficiency  is not
bounded by the standard Carnot limit.

\par
 According to above discussion, appearance of squeezed noise in future advanced devices is unavoidable, leading to destroy the equilibrium nature of the thermal state because of fast periodic modulations of the temperature. Thus, estimating the \textit{initial} temperature of  the  thermal bath driven by the squeezed noise is of great practical
importance.

 In this work, we  propose a thermometer, consisting N qubits for probing the initial temperature of a  thermal bath disturbed by the squeezed noise  and calculate  the bounds on the quantum thermometry in the presence of squeezing. It is supposed that $ n $ qubits are directly coupled to the
thermal bath of interest  
and $m~ (m=N-n) $ qubits  are not directly
coupled to the bath but instead serve as an information storage which may be read out at the final time $ t $. Our scheme 
relies on  the possibility of performing joint
measurements on all of $ N $ qubits. The model is well adopted to describe physical systems
such as the molecular oscillation, exciton-phonon interaction,
and photosynthesis process \cite{Dong1,Dong2,May2000}. Analytically, we investigate the effects of the initial state or squeezing and other environmental parameters on the estimation of the temperature. Besides, we extend our study to multiqubit estimation realized by initially entangled probes and address the role of entanglement in the process of thermometry  on the  squeezed thermal bath.

\par
\par This paper is organized as follows: In Secion \ref{pre}, we
present a brief review of the QFI and obtain the upper bound for the quantum  parameter estimation. The model is introduced in Section \ref{Model}. Different scenarios for quantum thermometry are discussed completely in Section \ref{st}.   Finally in Section \ref{conclusion}, the main results are summarized.

\section{The Preliminaries}\label{pre}

\subsection{(Quantum) Fisher information}
The classical Fisher information is an important method of measuring the amount of
information which an observable random variable $ X $ carries
about unknown parameter $ T $. Supposing that $ \{P_{i}(T)\}_{i=1}^{N} $ denotes the probability distribution  with measurement outcomes $ \{x_{i}\} $, the
classical Fisher information is defined as \cite{Helstrom1976}:

          \begin{equation}\label{FI}
\text{FI}_{T}=\sum_{i}P_{i}(T)\bigg(\dfrac{\partial~ \text{ln}P_{i}(T)}{\partial T}\bigg)^{2}
          \end{equation}
 characterizing the inverse variance of the asymptotic
normality of a maximum-likelihood estimator.  If observable $ \hat{X} $ is
continuous, the summation  should be replaced by
an integral.

 \par
 The quantum analog of the Fisher information can be formally
 generalized from Eq. (\ref{FI}) such that it is defined as \cite{V. GiovannettiPRL}:

          \begin{equation}\label{QFImohem}
F_{T}=\text{Tr}[\rho_{T} L_{T}^{2}]=\text{Tr}[\big(\partial_{T}~\rho_{T}\big) L_{T}],
          \end{equation}

in which the symmetric logarithmic derivative (SLD)
operator $ L_{T} $ represents   a Hermitian operator determined by

           \begin{equation}\label{SLD}
\partial_{T}\rho_{T}=\frac{1}{2}\{\rho_{T},L_{T}\},
           \end{equation}
  where  $ \{...\} $ denotes the anti-commutator. Considering the  density matrix \textit{spectral decomposition}   as  $\rho_{T}= \sum\limits_{i}\varrho_{i}|\psi_{i}\rangle\langle \psi_{i}| $, associated with $ \varrho_{i}\geq 0 $ as well as  $ \sum\limits_{i}\varrho_{i}=1 $,
  and focusing on the following formula for the QFI:
  \begin{equation}\label{qfiasli}
    F_{T}=2\sum\limits_{i,j} \dfrac{|\langle  \psi_{i}| \partial _{T}\rho_{T}| \psi_{j}\rangle|^{2}}{\varrho_{i}+\varrho_{j}},
  \end{equation}
   we can rewrite the QFI as \cite{Wei Zhong}
  \begin{equation}\label{aslqfi}
  F_{T}=\sum\limits_{i}\dfrac{(\partial_{T}\varrho_{i})^{2}}{\varrho_{i}} +2\sum\limits_{i\neq j} \dfrac{(\varrho_{i}-\varrho_{j})^{2}}{\varrho_{i}+\varrho_{j}}|\langle  \psi_{i}| \partial _{T} \psi_{j}\rangle|^{2}.
  \end{equation}
  where in the first and  second summations we should exclude 
   sums over
  all $ \varrho_{i}=0 $ and  $ \varrho_{i}+\varrho_{j}=0 $, respectively. 

According to quantum Cram\'{e}r-Rao (QCR) theorem, a significant property of the QFI is that we can obtain the
achievable lower bound of the mean-square error of the
unbiased estimator for  parameter T, i.e., 

          \begin{equation}
\text{Var}(\hat{T})\geq \dfrac{1}{MF_{T}},
          \end{equation}
in which  $ \hat{T} $ denotes the unbiased
estimator, and $ M $ represents the number of repeated
experiments.

  \subsection{Upper bound for parameter estimation}
  
  Given an initial pure state $\rho _{0}=|\psi_{0}\rangle \langle \psi_{0}| $, we know that it  evolves according to the expression $ \rho_{T}=\sum \varPi_{l}(T)\rho_{0} \varPi_{l}^{\dagger}(T) $, where  $\varPi_{l}(T) $'s represent $ T $-dependent Kraus operators \cite{Nielson}. It has been derived that the
  upper bound to the QFI is given by \cite{Eschernature}:

    \begin{equation} \label{Upper}
  C_{T}(\rho_{0},\varPi_{l}(T))=4[\langle I_{1}\rangle-\langle I_{2}\rangle^{2}]
    \end{equation}
  where
  
      \begin{equation} \label{Upper1}
    I_{1}(T)=\sum\limits_{l}^{}\frac{d\varPi^{\dagger}_{l}(T) }{dT}\frac{d\varPi_{l}(T) }{dT},
      \end{equation}

       \begin{equation} \label{Upper2}
      I_{2}(T)=i\sum\limits_{l}^{}\frac{d\varPi^{\dagger}_{l}(T) }{dT}\varPi_{l}(T),
        \end{equation}
  and $ \langle x \rangle\equiv\text{Tr}(x \rho_{0}) $.

\section{The Model   \label{Model}}
At first, we introduce the dephasing model \cite{Breuer}, 
composed of a two level system $ (H_{S}=\frac{1}{2}\Omega_{0}\sigma_{z})$ interacting with a boson bath
$ (H_{B}=\sum_{k}\omega_{k}b^{\dagger}_{k}b_{k}) $. The interaction of system-bath (S-B) can be
described by Hamiltonian $ V_{SB}=\sigma_{z}.\bigg\lgroup \sum_{k}(g_{k}b_{k}+g^{*}_{k}b^{\dagger}_{k})\bigg\rgroup $,
where  $ |g\rangle $ and $ |e\rangle $  are the ground and excited 
 states, respectively.
Because  $ [H_{S},V_{SB}]=0 $; the  energy of the system is 
conserved, and hence the population  $ p_{e},p_{g}$ of energy levels $ |e\rangle,|g\rangle$  do not change with
time.  
 \par

Starting from $ \rho_{SB}(0)=\rho_{S}(0)\otimes \rho_{B}(0) $, we focus on the case in which the initial state of the boson
bath is a squeezed thermal state \cite{Banerjee2007}

       \begin{equation} \label{Squeez}
    \rho_{B}(0)=S\rho_{th}S^{\dagger},~~~~\rho_{th}=\dfrac{1}{Z}\text{e}^{-H_{B}/T},
        \end{equation}
in which T represents the temperature of the thermal bath prepared in the state $ \rho_{th} $, $ Z $ denotes the
normalization constant, and $ S=\prod\limits_{k}^{}s_{k}$ represents the squeezing operator
for the boson bath, where $ s_{k} $ denotes the squeezing operator corresponding to  $ b_{k} $
mode \cite{Scully}:
       \begin{equation} \label{SqueezingO}
   s_{k}=\text{exp}\big(\frac{1}{2}\xi^{*}_{k}b^{2}_{k}-\frac{1}{2}\xi_{k}(b^{\dagger}_{k})^{2}\big),~~~~\xi_{k}=r_{k}\text{e}^{i\theta_{k}},
        \end{equation}
where $ r_{k} $ and $ \theta_{k} $ represent the squeezing strength and phase parameters,
respectively.
\par
In the interaction picture, the evolved reduced density matrix can be obtained as
\begin{equation}\label{Reduced}
\rho_{S}(t)=\left(\begin{array}{cc}
       p& q~\text{e}^{-\Gamma(t)}  \\
       q^{*}~\text{e}^{-\Gamma(t)} & 1-p \\
      \end{array}\right) 
      \end{equation}
      where $ p\equiv p_{e} $ is the population of excited level $ |e\rangle $ and $ \Gamma(t) $ denotes the decay factor. We focus on the Ohmic environment  where its coupling spectral density $J(\omega)=2\pi \sum\limits_{k}^{}|g_{k}|^{2} \delta(\omega-\omega_{k}) $,  whose summation
      should be written as an integral for continuous bath modes, is given by 
      $J(\omega)=\lambda \omega \text{e}^{-\omega/\Omega_{c}}  $ in which $ \Omega_{c} $ denotes the cutoff frequency and $ \lambda $ is a unitless number representing the
      coupling strength.
      \par
      Defining $ \varDelta\theta_{k}\equiv \theta_{k}-2\phi_{k} $ where $ \phi_{k}=\text{arg}[g_{k}] $ and assuming that $ r_{k}=r $ and $\varDelta\theta_{k}=\delta \theta  $,  we can find that the decay factor is given by \cite{YiNingYou} 
       
            \begin{equation} \label{Gamma}
        \Gamma(t)= \frac {\lambda}{\pi } \bigg( A_{{t}}\cosh 2r -\sinh
         2r  \left( B_{{t}}\cos \left( \delta \theta \right) +C_{{t}
         }\sin \left( \delta \theta \right)  \right)  \bigg)
               \end{equation}
where, as described before, $ r $ represents the squeezing
strength, $ \lambda $  characterizes the S-B coupling
strength, and $\delta \theta  $ denotes the phase difference between the squeezing
phase $ \theta $ relative to the phase of the coupling strength. Moreover, the  time-dependent coefficients are of the form

     \begin{equation}\label{key1}
A_{t}=a_{\eta}+\sum\limits_{n=1}^{\infty}Ta_{t}(n),\\~~
    B_{t}=b_{\eta}+\sum\limits_{n=1}^{\infty}Tb_{t}(n),\\~~
         C_{t}=c_{\eta}+\sum\limits_{n=1}^{\infty}Tc_{t}(n),\\
            \end{equation}
where
\begin{equation}
a_{\eta}=\ln  \left( {\eta}^{2}+1 \right) ,\\~~
b_{\eta}=\ln  \left( {\frac {\sqrt {4\,{\eta}^{2}+1}}{{\eta}^{2}+1}} \right) ,\\~~
c_{\eta}=2\,\text{tan}^{-1} \left(\eta \right) -\text{tan}^{-1} \left( 2\,\eta\right) ,\\\nonumber
\end{equation}
     and  where 
     \begin{equation}
     a_{t}=\frac {2t }{\tau} \left( 2\,\tau\,\text{tan}^{-1} \left( \tau \right) -\ln 
      \left( {\tau}^{2}+1 \right)  \right),\\
       \end{equation}
            \begin{equation}
     b_{t}= \frac {2t}{\tau} \left( 2\,\tau\, \left( \text{tan}^{-1} \left( 2\,\tau
      \right) -\text{tan}^{-1} \left( \tau \right)  \right) -\ln  \left( {\frac {
      \sqrt{4\,{\tau}^{2}+1}}{{\tau}^{2}+1}} \right)  \right),\\
      \end{equation}
      \begin{equation}
     c_{t}=\frac {2t}{\tau} \left( \tau\,\ln  \left( {\frac {4\,{\tau}^{2}+1
     }{{\tau}^{2}+1}} \right) -2\,\text{tan}^{-1} \left( \tau \right) +\text{tan}^{-1}
      \left( 2\,\tau \right)  \right) 
      ,\\\nonumber
     \end{equation}  
     in which $ \eta=\Omega_{c}t $ and $\tau=\big\lgroup\Omega_{c}/\big(1+\dfrac{n\Omega_{c}}{2T}\big)\big\rgroup t $.
     
\section{Scenarios for quantum thermometry   \label{St}}
  \begin{figure}[ht!]
                                   \subfigure[]{\includegraphics[width=7.1cm]{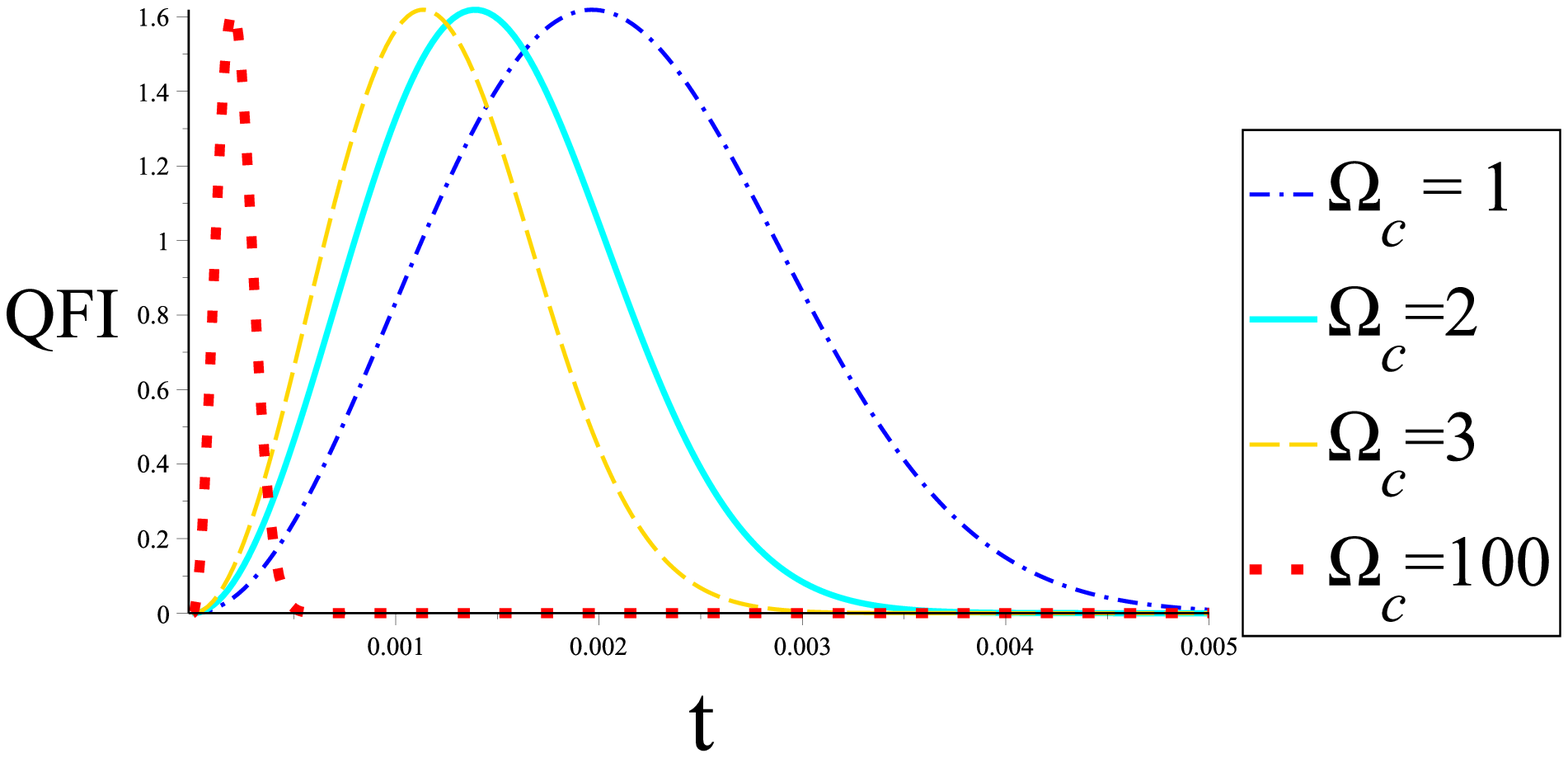}\label{qfi1omega}}
                                   \subfigure[]{\includegraphics[width=7.1cm]{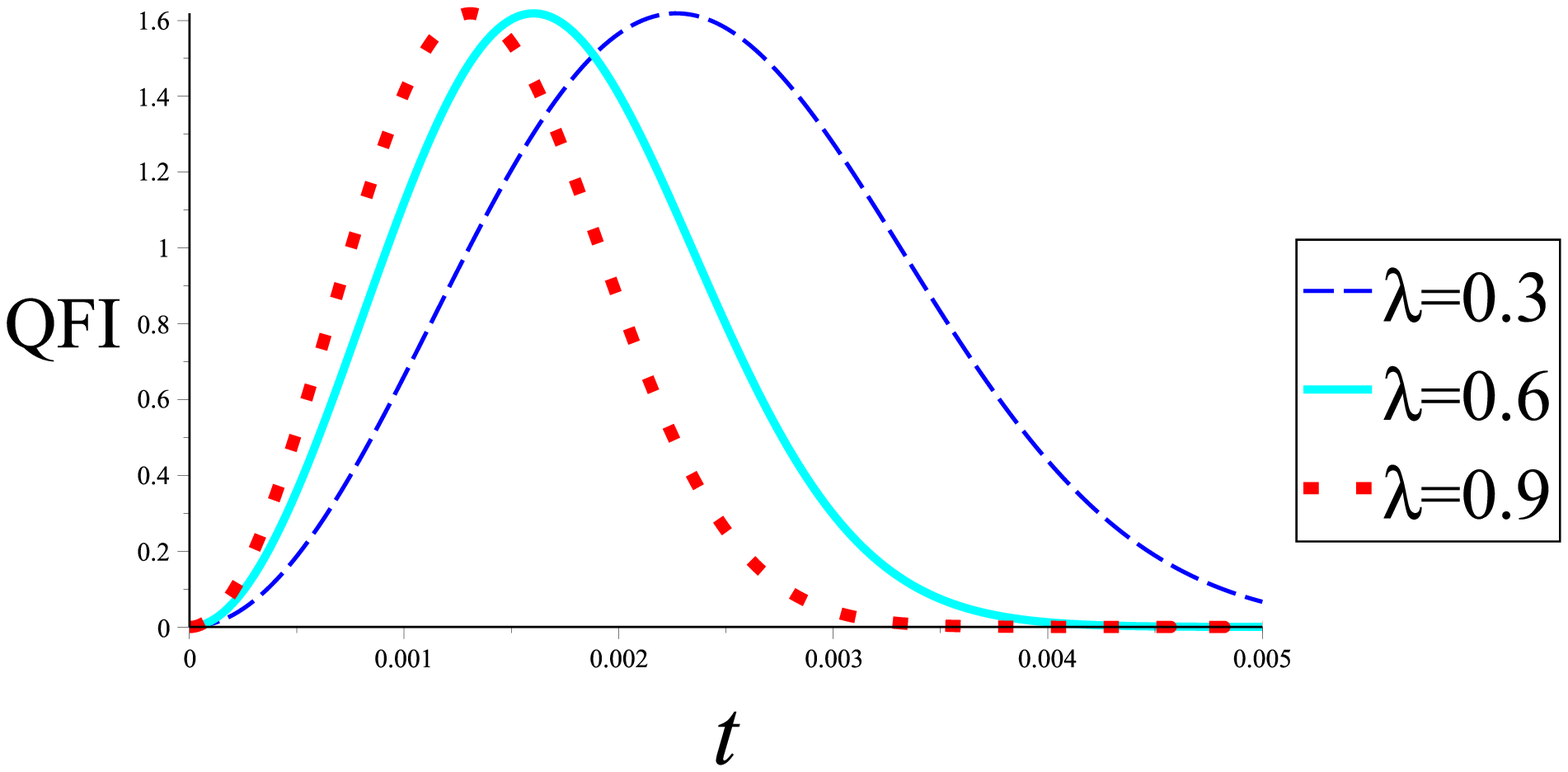}\label{qfi1lambda}}
                                       \caption{\small  (a) Time variation of the normalized single-qubit QFI associated with the temperature for $ \lambda = 0.4$, $r = 0.5$,  $\delta \theta = 0.9 $ and different  cutoff frequencies. (b) The same quantity for  $ r = 0.5 $, $ \delta\theta=0.9 $ and different values of S-B coupling strength.}
                                                                        \label{qfi1omlam}
     \end{figure}
        \begin{figure}[ht!]
                                             \subfigure[]{\includegraphics[width=7.1cm]{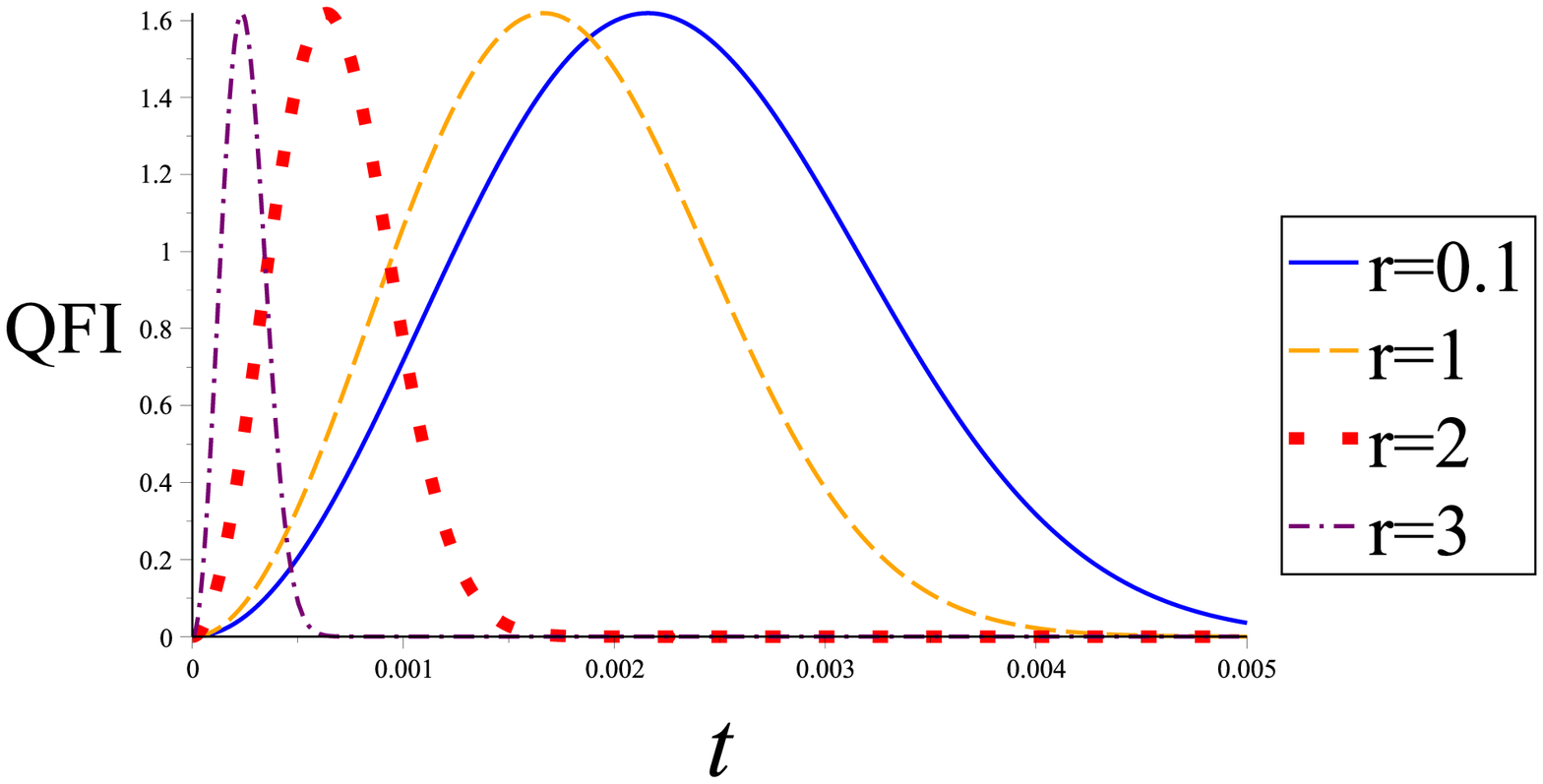}\label{qfi1r1}}
                                             \subfigure[]{\includegraphics[width=7.1cm]{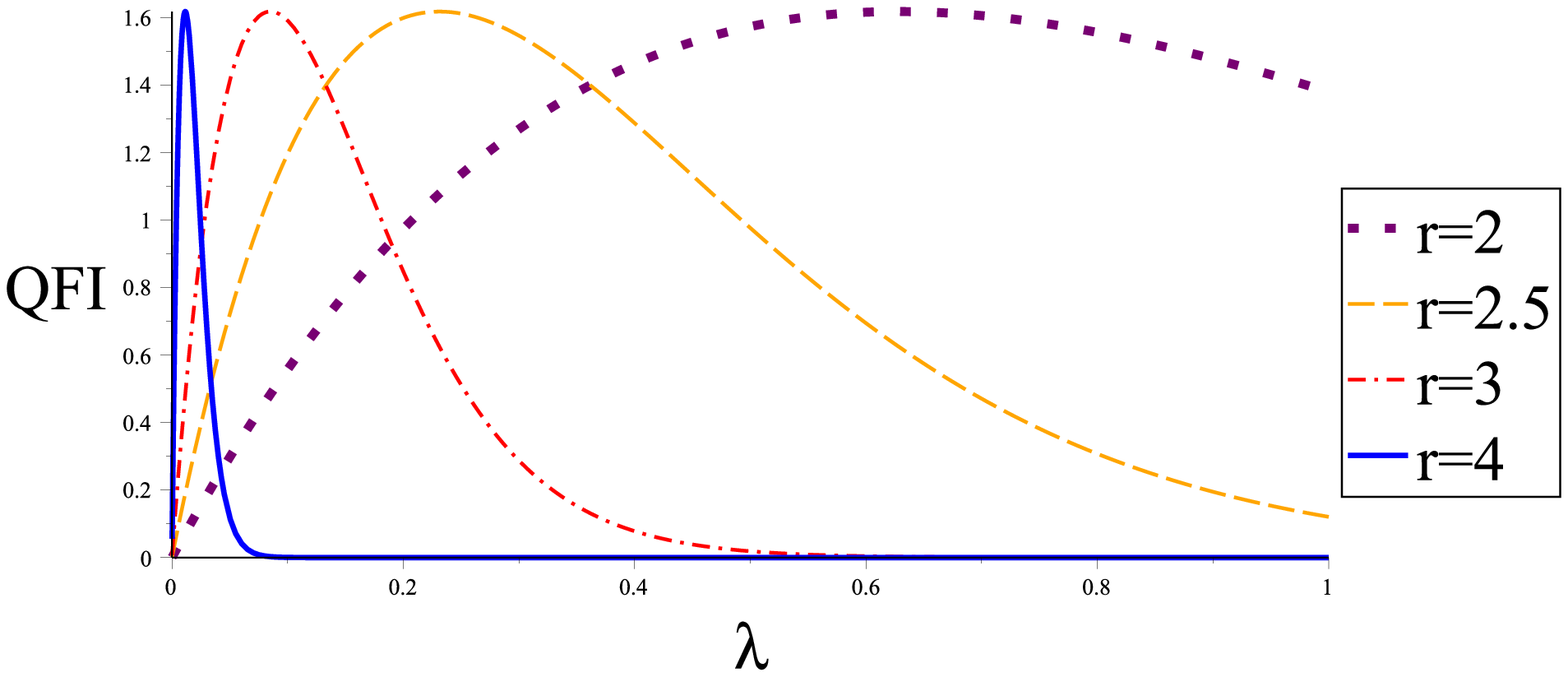}\label{qfi1r2}}
                                                 \caption{\small  (a) Dynamics of normalized single-qubit QFI associated with estimating $ T $ for $ \lambda = 0.3$,  $\delta \theta = 0.9 $ and  different values of the squeezing strength. (b) The same quantity versus S-B coupling strength   for  $ \Omega_{c}t= 0.0005 $, $ \delta\theta=5.5 $ and different values of $ r $.}
                                                                                  \label{qfi1r12}
               \end{figure}
                         \begin{figure}[ht!]
                                                                 \subfigure[]{\includegraphics[width=7.1cm]{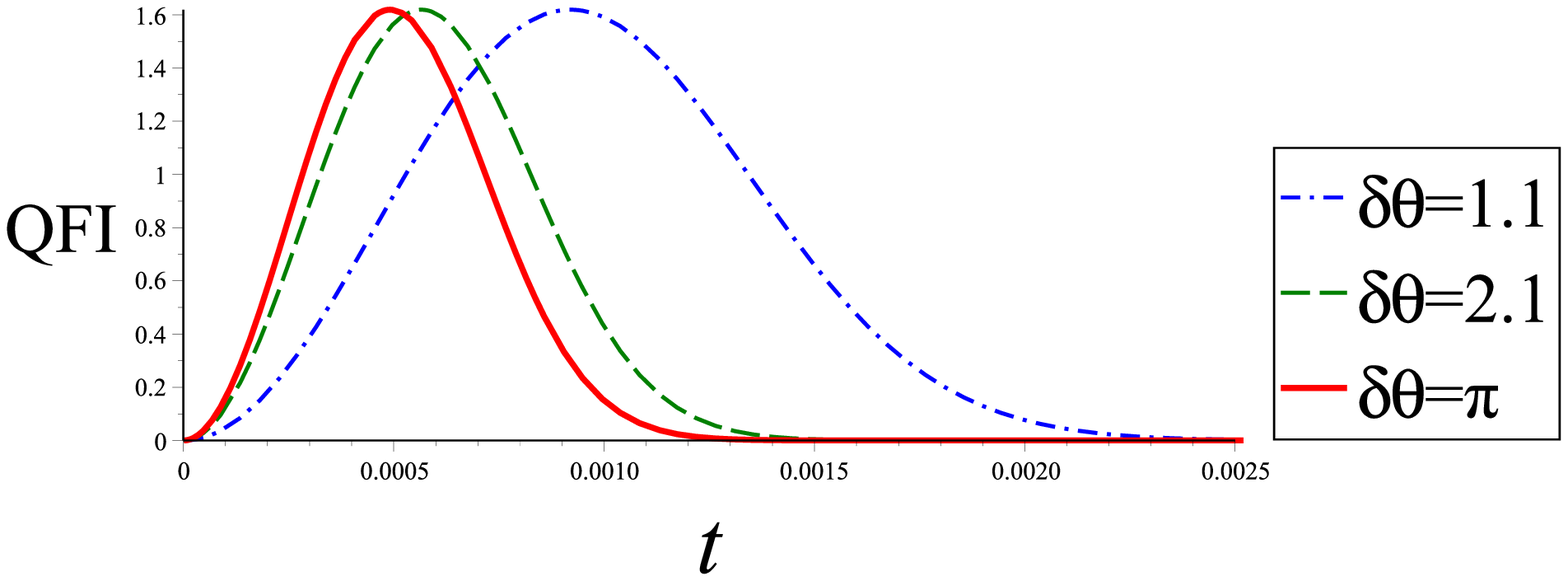}\label{qfi1delta1}}
                                                                 \subfigure[]{\includegraphics[width=7.1cm]{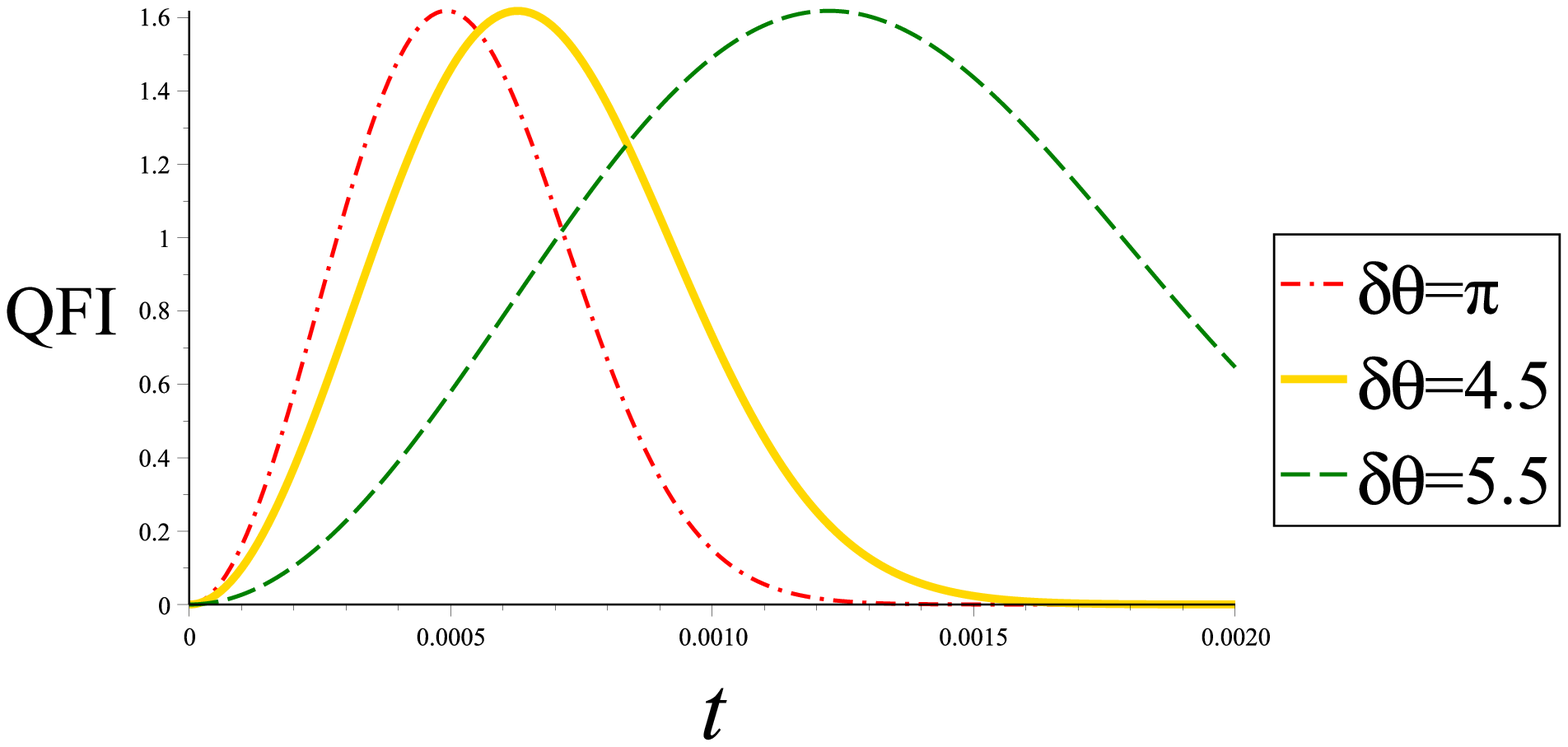}\label{qfi1delta2}}
                                                                     \caption{\small  Dynamics of the normalized single-qubit QFI for $ \lambda=0.7 $, $ r=1 $, and different values of the squeezing phase $ \delta \theta $ lying in the regions (a) $ [\frac{\pi}{4},\pi] $, and (b) $ [\pi,2\pi] $.  }
                                                                                                      \label{qfi1delta12}
                                   \end{figure}
     \subsection{Single-qubit scenario   \label{stl}}
     
     Our model describes a dephasing channel $ \mathcal{E}_{T} $ such that the system can be used for probing  temperature $ T $ of the thermal bath. In the first scenario in which the single-qubit is used for thermometry, the corresponding QFI is obtained as follows:
     
   \begin{equation}\label{QFI1Q}
    F_{T} =  \frac {4 \left( p-1 \right) p   \left| q \right|  ^{2
       }}{   \left| q \right|  ^{2}+{{\rm e}^{2\,\Gamma }}
        \left( p-1 \right) p}\big(\dfrac{\partial \Gamma}{\partial T}\big)^{2}
        \end{equation}    
     where the partial derivative is given by
     
     \begin{equation} \label{DGamma}
             \dfrac{\partial \Gamma}{\partial T}= \frac {\lambda}{\pi } \bigg(\mathcal{A} _{{t}}\cosh 2r -\sinh
              2r  \left( \mathcal{B}_{{t}}\cos \left( \delta \theta \right) +\mathcal{C}_{{t}
              }\sin \left( \delta \theta \right)  \right)  \bigg)
                    \end{equation}
     in which 
     
     \begin{equation}
        \mathcal{A}_{t}= 2\,\sum _{n=1}^{\infty }\bigg\lgroup 4\,t~\text{tan}^{-1}  \left(\zeta \right) -\,{\frac {2}{\Omega_{{c}}}\ln  \left( 
         \zeta^{2}+1 \right) }\bigg\rgroup
         ,\\\nonumber
            \end{equation}
            \begin{equation}
              \mathcal{B}_{t}=  2\,\sum _{n=1}^{\infty }\biggl\{\,{\frac {-2}{\Omega_{{c}}}\ln  \left( {\frac {\sqrt {4\,{\zeta }^{2}
                          +1}}{{\zeta }^{2}+1}} \right) }-4\,t~ \text{tan}^{-1} \left( \zeta  \right) +4\,
                          t~\text{tan}^{-1} \left( 2\,\zeta  \right)\biggl\}
                     \end{equation}

           \begin{equation}
          \mathcal{C}_{t}=2 \sum _{n=1}^{\infty } \frac{2}{\text{$\Omega_{c} $}} \bigg\{t \text{$\Omega_{c} $} \ln \left( {\frac {4\,{\zeta }^{2}
                                    +1}{{\zeta }^{2}+1}} \right)-2 \tan ^{-1}\left(\zeta\right)+\tan ^{-1}\left(2\zeta\right)\bigg\}
           ,\\\nonumber
          \end{equation}  
     where $ \zeta\equiv\,{\frac {2tT\Omega_{{c}}}{n\Omega_{{c}}+2\,T}} $.

     \par
 Preparing the qubit probe  in a pure state   $ |\psi_{0}\rangle=\text{cos}(\theta_{0}/2)|0\rangle+\text{sin}(\theta_{0}/2)\text{e}^{i\varphi_{0}}|0\rangle $, the QFI reduces to the following expression:
   \begin{equation}\label{QFI1Qp}
      F_{T} =\frac{1}{2}\big(\coth(\Gamma)-1\big)\sin^{2}(\theta_{0})  \big(\dfrac{\partial \Gamma}{\partial T}\big)^{2},
          \end{equation}    
  saturating the upper bound obtained from Eq. (\ref{Upper}) for $ \theta_{0}=\pi/2 $. Throughout this paper, we set $ \theta_{0}=\pi/2 $ and normalize the QFI for clearer illustration.
  
       \par
       
 Although our approach for computing the  analytical results
        is completely general, we
  limit our study of the QFI behaviour to the high-T Ohmic reservoir. Therefore, all QFI figures are plotted and interpreted in this regime. In the high-temperature limit that   $ \text{coth}(\omega/2T)\approx 2T/\omega $ and $ \tau \approx \eta=\Omega_{c}t $, it is found that
       
         \begin{equation}
  \mathcal{A}_{t}\approx T a_{t},~~~\mathcal{B}_{t} \approx T b_{t},~~~\mathcal{C}_{t} \approx T c_{t}.
                \end{equation}

     \par

  Investigating the QFI as a function of time, we find that  at first it increases, reaching  a maximum (see Figs. \ref{qfi1omlam}-\ref{qfi1delta12}).
                      According to theory of quantum metrology, an increase
                      in QFI means that the  precision of quantum estimation
                      is improved. This originates from the fact that interaction of the qubit with the bath encodes the information about the temperature  into the quantum state of
                      the probe, hence the QFI increases. However,
                      because of the decoherence effects, the encoded information flows from the system to the environment, and hence its destructive influence appears and the QFI falls, thus, the
                           quantum thermometry becomes more inaccurate.

                          We first  investigate the behaviour of the QFI dynamics with respect to cutoff frequency $ \Omega_{c} $ and S-B coupling strength $ \lambda $. As shown in Figs. \ref{qfi1omega} and \ref{qfi1lambda}, the variation of theses  parameters can not change 
                               the maximum value of the QFI. Therefore, the optimum  precision  obtained  in the process of thermometry does not vary. However, the figure shows when the cutoff frequency or the coupling increases,
                                  the maximum point of the QFI is shifted to the
                                 left, and hence the QFI reaches its maximum value sooner. Although increase of the coupling  between the probe and the bath or increasing the cutoff frequency decreases the interaction time for obtaining the optimal
                                 estimation precision, the QFI decreases more quickly, and hence the time period that we can extract the information about the temperature is shortened.

                                 Squeezing effects of increasing squeezing strength $ r $ are shown in Fig. \ref{qfi1r12}. We can achieve sooner the optimal precision of thermometry probing more squeezed fields such that the optimal value of the QFI also remains invariant. Interestingly, Fig. \ref{qfi1r2} illustrates that in more squeezed fields the optimal estimation is attainable with more weak coupling between the probe and the bath. This may leads to important results in improving the control of decoherence in the process of quantum communication.

Now we investigate how $ \delta\theta $ may affect the dynamics of the QFI. As shown in Fig. \ref{qfi1delta12}, when the relative phase varies from $ \frac{\pi}{4} $ to $ \pi $ \text{rad}, the interaction time between the probe and the bath can be reduced and hence the optimal value of estimation, not affected by variation of the relative phase, is obtained sooner.  However, varying the relative phase from $ \pi$ to $ 2\pi $, we see that  the maximum point, at which the QFI is  maximized, can be shifted to the
right. Therefore, the QFI reaches its maximum value at a
later time-point. Nevertheless, Fig. \ref{qfi1delta2} exhibits a positive and
interesting  consequence of increasing the relative phase from $ \pi$ to $ 2\pi $ . We see that  larger relative phases lead to retardation of 
 the QFI
loss during the time evolution and therefore enhance the estimation of the parameter at a later time.

       \subsubsection{ Feasible measurement for optimal estimation   \label{PI}}
       Another important question is how we can practically design
       the optimal estimation, i.e., a practically feasible measurement  whose Fisher information is equal to the QFI. For answering this question, we should compute the SLD, since the optimal POVM can be constructed by
       the eigenvectors of the SLD \cite{M.G.A. Paris,RanganiOPTC}. Using Eq. (\ref{Reduced}) and following the approach introduced in 
       \cite{Chapeau2015} for computing the SLD of one-qubit systems,
        we find that the SLD associated with the thermometry
       is given by
       
       \begin{equation}\label{SLDasli}
      L_{T}=(\frac{\partial \Gamma}{\partial T})^{2} \left( \begin {array}{cc} {\frac {2 \left( p-1 \right) \left| q \right|   ^{2}}{   \left| q \right|   ^{2}+{{\rm e}^{2\Gamma}}
        \left( p-1 \right) p}}&\,{\frac {-2{\rm e}^{\Gamma} \left( p-1 \right) p
       q}{   \left| q \right|   ^{2}+{\rm e}^{2\Gamma} \left( p-1
        \right) p}}\\ \noalign{\medskip}{\frac {{-2{\rm e}^{\Gamma}} \left( p-1
        \right) p\overline{q}}{   \left| q \right|   ^{2}+{
       {\rm e}^{2\Gamma}} \left( p-1 \right) p}}&{\frac {-2p\left| q \right|   ^{2}}{
          \left| q \right|   ^{2}+{{\rm e}^{2\Gamma}} \left( p-1
        \right) p}}\end {array} \right).
       \end{equation}
      When the qubit probe is  prepared initially in a pure state $ |\psi_{0}\rangle=\text{cos}\dfrac{\theta}{2}|0\rangle+\text{sin}\dfrac{\theta}{2}\text{e}^{-i\varphi}|1\rangle $, some interesting results may be extracted. In particular, for $ \theta=\frac{\pi}{2} $, the SLD reduces to the following compact form:
       \begin{equation}
            L^{\theta=\frac{\pi}{2}}_{T}=(\frac{\partial \Gamma}{\partial T})^{2} \left( \begin {array}{cc} \dfrac{1}{\text{e}^{2\Gamma}-1}&\dfrac{1+\text{e}^{\Gamma+i\varphi}}{1-\text{e}^{2\Gamma}}\\ \noalign{\medskip} \dfrac{1+\text{e}^{\Gamma-i\varphi}}{1-\text{e}^{2\Gamma}}&\dfrac{1}{\text{e}^{2\Gamma}-1} \end {array} \right).
             \end{equation}
    For $ \varphi=0 $ ($ \varphi=\pi /2$ ), the above SLD  commutes with $ \sigma_{x}~ (\sigma_{y}) $ and they have common eigenvectors, i.e., measurement of $ \sigma_{x}~ (\sigma_{y}) $ leads to the optimal estimation of the temperature, because an optimal
    measurement can be performed if we measure in the eigenbasis of the SLD \cite{Geze}.
     \subsection{Multi-qubit scenario   }
     For our dephasing model, it is simple to obtain the operator-sum representation 
               $\rho_{S}(t)\equiv\mathcal{E}_{T}(\rho_{S}(0))=\sum\limits_{i}^{}k_{i}\rho_{S}(0)k^{\dagger}_{i}  $  where the time-dependent Kraus operators are given by 
               
          \begin{equation}\label{KR}
          k_{1}=\sqrt{\dfrac{1+\text{e}^{-\Gamma(t)}}{2}}\left(\begin{array}{cc}
                 1&0  \\
                 0& 1 \\
                \end{array}\right),~~~k_{2}=\sqrt{\dfrac{1-\text{e}^{-\Gamma(t)}}{2}}\left(\begin{array}{cc}
                       1&0  \\
                       0& -1 \\
                      \end{array}\right).  
                \end{equation}
                \par
\begin{figure}[ht!]
                                 \subfigure[]{\includegraphics[width=5cm]{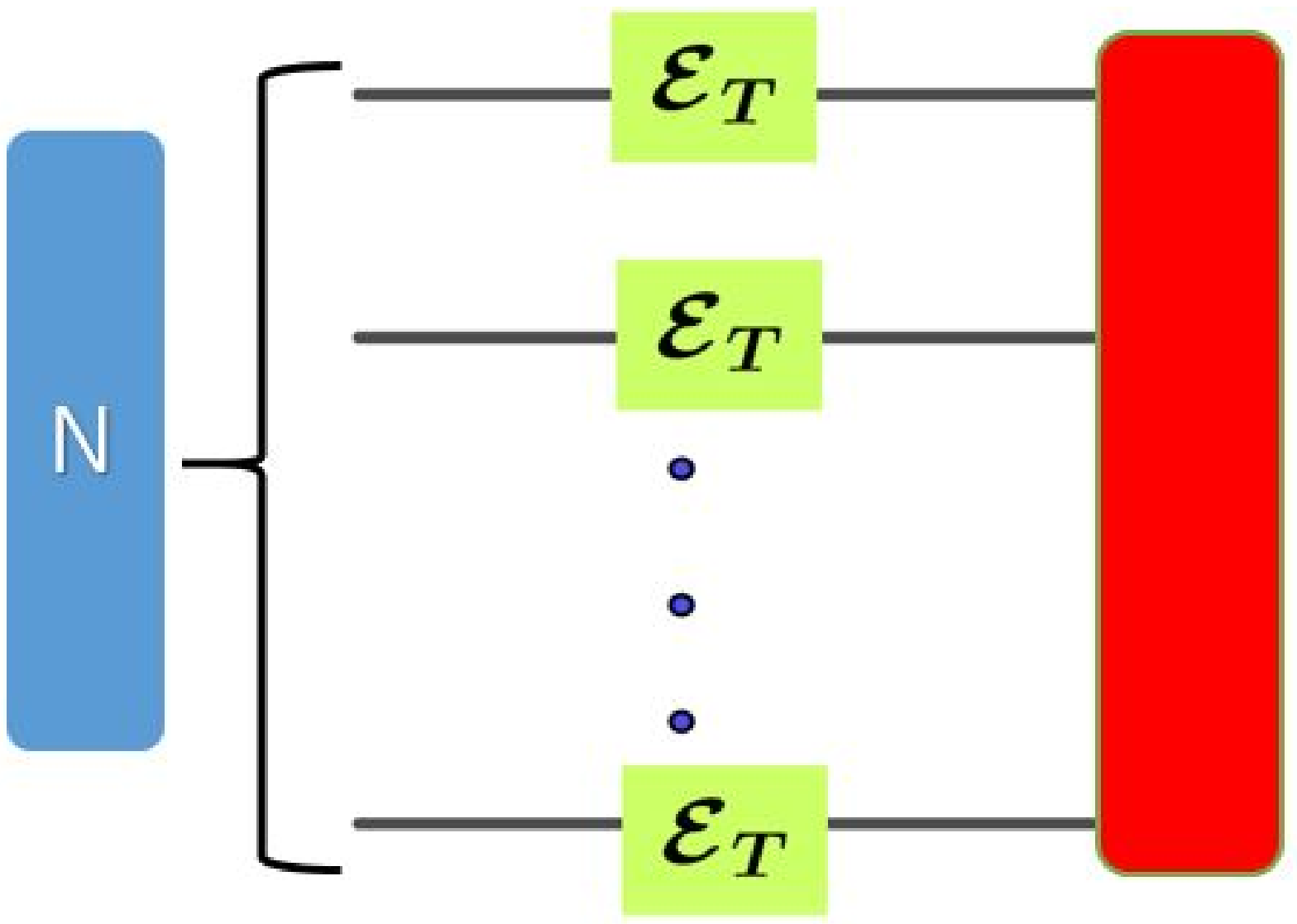}\label{st1}}
                                 \subfigure[]{\includegraphics[width=4cm]{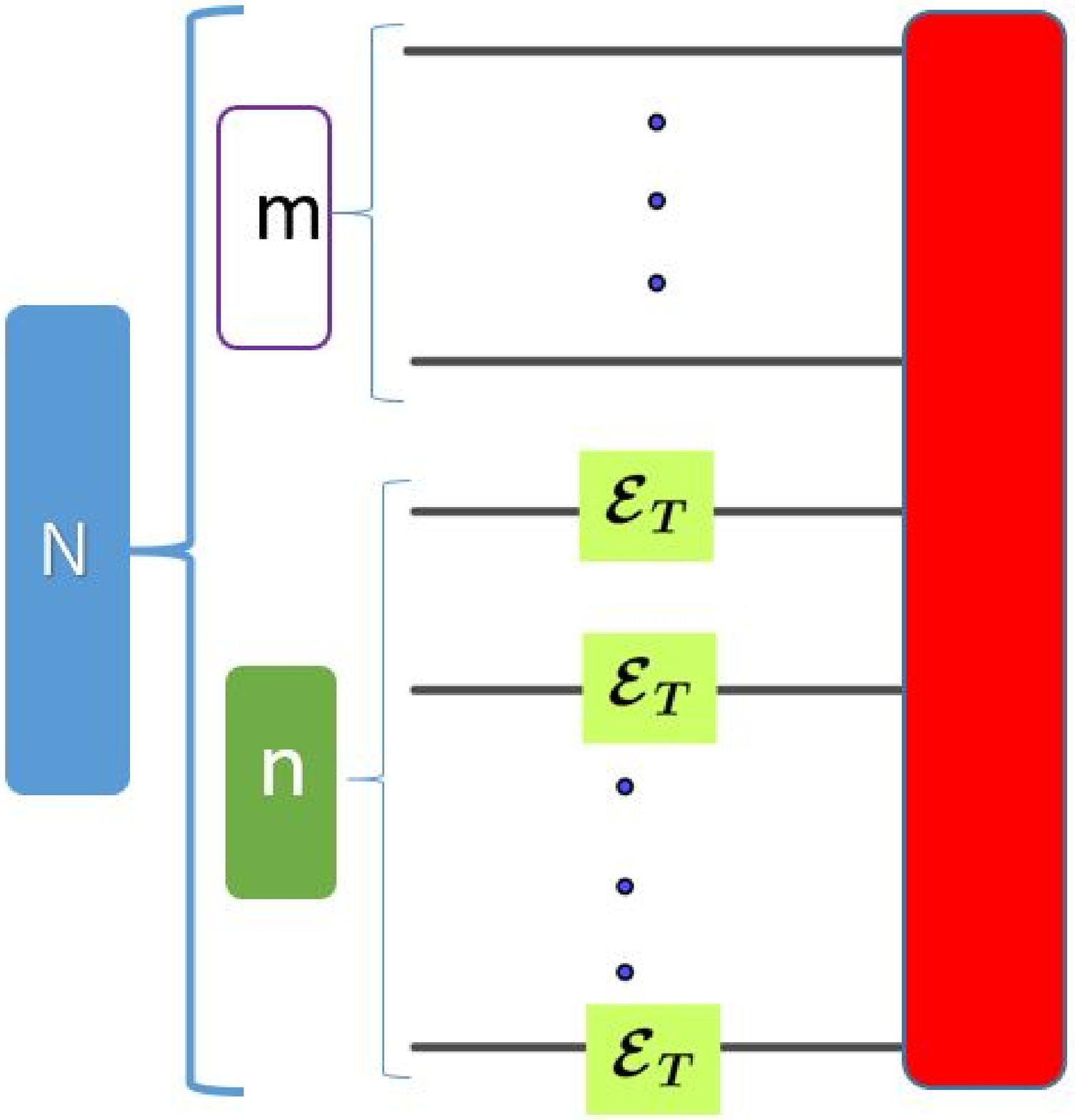}\label{st2}}
                                     \caption{\small   Two different quantum metrology strategies  (a) The parallel 
                                     strategy: a state of $ N $ (non-) entangled probes goes through $ N $ maps in parallel. (b) The ancilla-assisted
                                     strategy, in which $ n $ individual  probes entangled with  $ m $ noiseless ancillas
                                     go through the map.}
                                                                      \label{st}
   \end{figure}

      Two usual scenarios in which
     $ N $ probes are submitted to  \textit{independent} processes  are shown in Fig. \ref{st} for estimating  the temperature.   In both parallel   and ancilla-assisted strategies shown in Figs. \ref{st1} and \ref{st2},  using the above Kraus operators,
     we find that $ I_{1} $ and $ I_{2} $ introduced, respectively, in Eqs. (\ref{Upper1}) and (\ref{Upper2})   are given by:
     
     \begin{equation}
               I_{1}(T)=\frac{n}{8}\big(\coth(\Gamma)-1\big)  \big(\dfrac{\partial \Gamma}{\partial T}\big)^{2}\textbf{I}_{N\times N},~~I_{2}(T)=\textbf{0}
                     \end{equation}
        where $ \textbf{I}_{N\times N} $ denotes $ N\times N $ identity operator acting on $ N $-dimensional Hilbert space. Inserting these equations into (\ref{Upper}),  one obtains the upper bound for the QFI associated with the temperature  as follows:

      \begin{equation}\label{QFIuppm}
           C^{n}_{T} =\frac{n}{2}\big(\coth(\Gamma)-1\big)  \big(\dfrac{\partial \Gamma}{\partial T}\big)^{2},
               \end{equation}    
     where in  the parallel  strategy, we put $ n\equiv N $. 
     Although the use of  the
     noiseless ancillas does not improve the upper bound for the QFI, we cannot necessarily conclude that the parallel strategy leads to more accurate estimation than the ancilla-assisted one. 
     We will come
     back to discuss this problem, after obtaining the exact expressions for the QFIs.
     \par

     In the first scenario, we adopt the ancilla-assisted strategy in the sense that the probes are realized with two qubits such that one of which is noiseless ( see Fig.  \ref{st2} with $ n=1 $, and $ m=1 $). 
     Because this is the model for two independent  environments, the Kraus operators are just tensor products of
     Kraus operators of each of the qubits, $ K_{1,2}=I\otimes k_{1,2} $. Preparing initially the qubits in  the Bell state $ |\psi_{Bell}\rangle=\dfrac{1}{\sqrt{2}}\big(|00\rangle+|11\rangle\big) $, we find that the output state of the channel,
       measured for estimating the temperature, is given by
      \begin{equation}\label{rhoout}
           \rho^{Bell}_{\text{out}}=\left(\begin{array}{cccc}
                  1/2&0&0&\text{e}^{-\Gamma}/2  \\
                  0&0&0&0  \\
                   0&0&0&0  \\
                    \text{e}^{-\Gamma}/2&0&0&1/2  \\
                 \end{array}\right).
                 \end{equation}
Therefore, the corresponding QFI denoted by $QFI_{N,n} $, is obtained as 

 \begin{equation}\label{Bell}
QFI_{2,1}(\rho^{Bell}_{\text{out}})=\frac{1}{2}\big(\coth(\Gamma)-1\big)  \big(\dfrac{\partial \Gamma}{\partial T}\big)^{2}\equiv C^{1}_{T},
                           \end{equation}
saturating the upper bound. In the parallel scenario in which both qubits are affected by the noise, the QFI is given by: 
 \begin{equation}\label{Bell2}
QFI_{2,2}(\rho^{Bell}_{\text{out}})=\dfrac{4}{\text{e}^{4\Gamma}-1} \big(\dfrac{\partial \Gamma}{\partial T}\big)^{2},
                           \end{equation}

The saturation of the upper bound in the ancilla-assisted strategy indicates that this scenario  may sometimes lead to more accurate estimation than the parallel one.
\par 
An important difference between bipartite entanglement and multipartite one is how they are classified. Greenberger-Horne-Zeilinger (GHZ) state \cite{Greenberger11311990}, W state \cite{Dur0203032001} are two typical classes of
multipartite entangled states needed for different quantum information
processing tasks. For instance, GHZ states are the best quantum
channels for teleportation \cite{Zhao542004}  or quantum key distribution \cite{Kempe9101999}, and W states are required for secure quantum communication \cite{Wang6372007,Liu31602011}. Moreover,  the entanglement of W state is robust against disposal of particles \cite{Dur0203032001}. Besides, the W state plays an important
role in the leader election problem in anonymous quantum networks \cite{DHondt1732006}. It has been also shown that the quantum coherence of W states leads to high efficiency in quantum thermalization of a single mode cavity \cite{Mustecaplioglu}.

 Using a three-qubit probe ($ N=3 $) with $ n=1 $, and $ m=2 $, initially prepared in the GHZ state $ |\psi_{GHZ}\rangle=\dfrac{1}{\sqrt{2}}\big(|000\rangle+|111\rangle\big) $, we again the same result for $ QFI_{2,1}(\rho^{Bell}_{\text{out}}) $, i.e.,

 \begin{equation}\label{GHZ3}
QFI_{3,1}(\rho^{GHZ}_{\text{out}})=QFI_{2,1}(\rho^{Bell}_{\text{out}}).
                           \end{equation}
On the other hand, if the probes are initially in the W state $ |\psi_{W}\rangle=\dfrac{1}{\sqrt{3}}\big(|001\rangle+|010\rangle+|100\rangle\big) $
the QFI is given by

 \begin{equation}\label{GHZ4}
QFI_{3,1}(\rho^{W}_{\text{out}})=\frac{4}{9}\big(\coth(\Gamma)-1\big)  \big(\dfrac{\partial \Gamma}{\partial T}\big)^{2}.
                           \end{equation}

 Therefore,  $ QFI_{3,1}(\rho^{W}_{\text{out}}) < QFI_{3,1}(\rho^{GHZ}_{\text{out}})  $ and it cannot saturate the upper bound.

Natural generalizations of the GHZ and W states to N-qubit systems
are
\begin{equation}\label{GHZN}
|GHZ\rangle_{N}=\dfrac{1}{\sqrt{2}}\big(|0\rangle^{\otimes N}+|1\rangle^{\otimes N}\big),
\end{equation}
\begin{equation}\label{WN}
|W\rangle_{N}=\dfrac{1}{\sqrt{N}}\big(|10000...0\rangle+|0100...0\rangle+|0010...0\rangle+...+|0000...1\rangle\big).
\end{equation}

 Starting from initial state $ |GHZ\rangle_{N} $ such that $ n $ qubits are affected by the channel and using Eq. (\ref{aslqfi})  for computing the QFI corresponding to the temperature estimation, we find that the second term in right hand side  of  (\ref{aslqfi}) is always zero because the eigenvectors of the evolved density matrix are $ T $-independent. Moreover, all the eigenvalues except  two  vanish, simplifying the computation of the QFI. Hence, simultaneously using  \textit{Mathematica} and \textit{QUBIT4MATLAB V5.6} \cite{TothCPC}  to  work with the high-dimensional density matrices, we find that the corresponding QFI for any  choice of $\{N,n\} $ is obtained as follows:

 \begin{equation}\label{QFIGHZN}
 QFI_{N,n}(\rho^{GHZ}_{\text{out}}) = \left\{
 \begin{array}{rl}
 \frac{1}{2}\big(\coth(\Gamma)-1\big)  \big(\dfrac{\partial \Gamma}{\partial T}\big)^{2}\equiv F^{\theta_{0}=\pi/2}_{T} \equiv C^{1}_{T}  &~~~~~~~~  n=1\\
 ~~\\
 \dfrac{n^{2}}{\text{e}^{2n\Gamma}-1}  \big(\dfrac{\partial \Gamma}{\partial T}\big)^{2} & ~~~~~~~~ n> 1.
 \end{array} \right.
 \end{equation}
  Clearly, an increase in the number of the noiseless qubits does not affect the accuracy of the estimation. The above formula can be generalized to parallel strategy as follows:
 
 \begin{equation}\label{QFIGHZNPAR}
  QFI_{N,N}(\rho^{GHZ}_{\text{out}})=\dfrac{N^{2}}{\text{e}^{2N\Gamma}-1}  \big(\dfrac{\partial \Gamma}{\partial T}\big)^{2}~~~~~~\text{for}~~~N\geq 3. 
  \end{equation}

     \begin{figure}[ht!]
              \subfigure[]{\includegraphics[width=7.1cm]{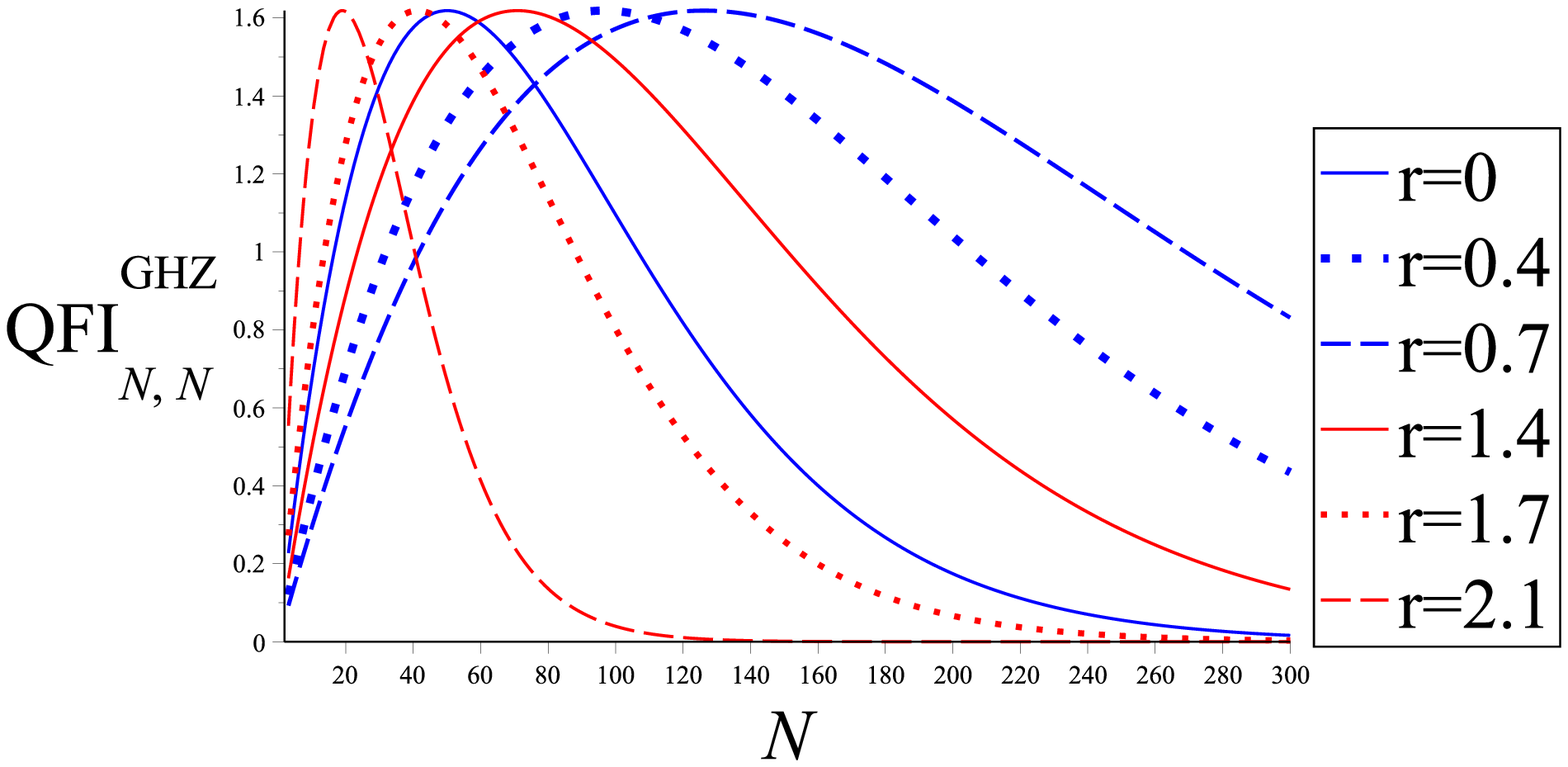}\label{qfighz1}}
            \subfigure[]{\includegraphics[width=7.1cm]{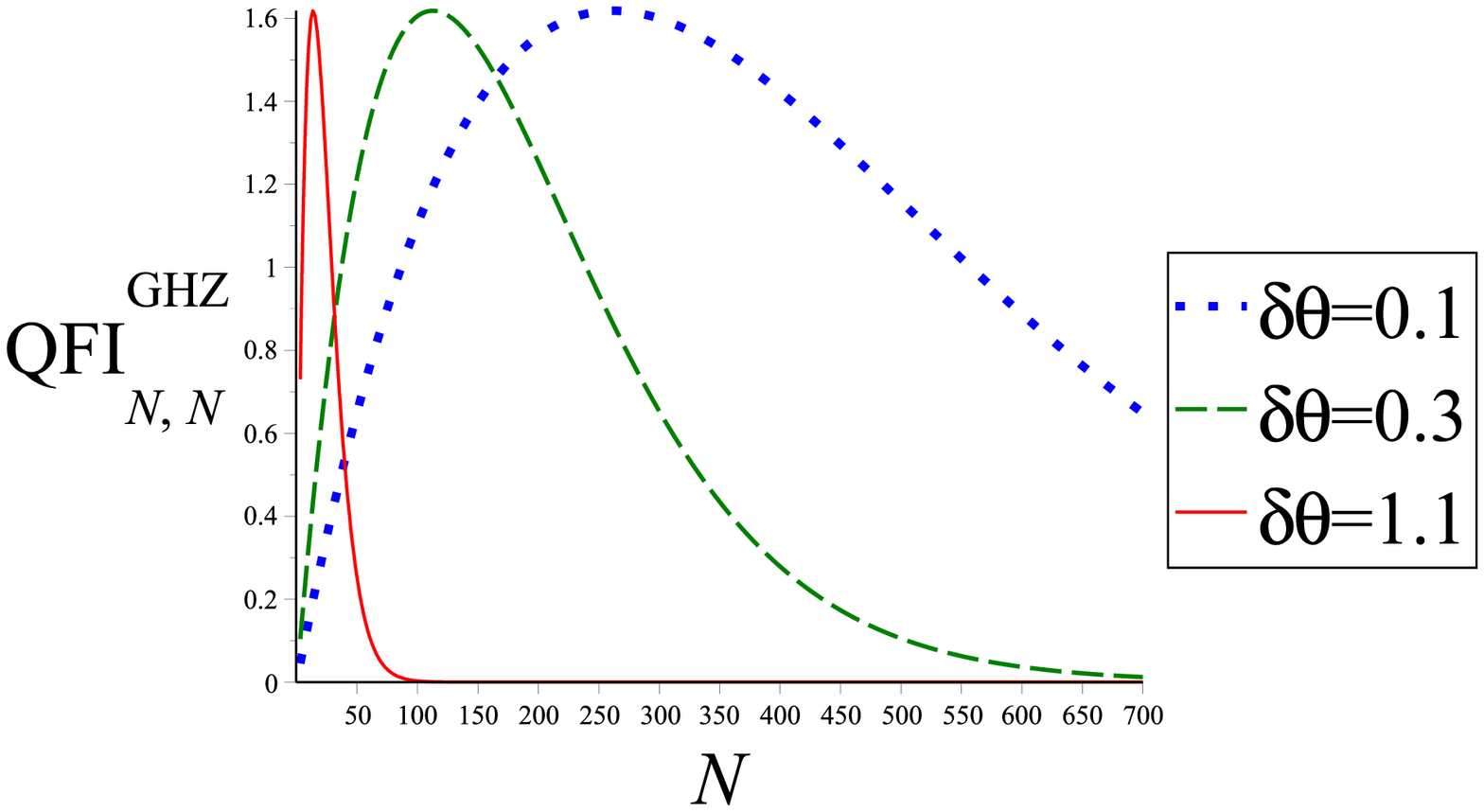}\label{qfighz2}}
                            \caption{\small Effects of squeezing parameters on the normalized multi-qubit QFI   with respect to the number of uses of the channel in the parallel strategy, starting from GHZ state, for $ \lambda = 0.1 $,  (a) $ \delta\theta = 0.4, t = 0.0005  $,   (b)  $ r =1.1$ and $t = 0.0006 $.}
                                                                                                         \label{qfighz12}
    \end{figure}

As seen in Fig. \ref{qfighz12}, we find that in the parallel strategy with initial GHZ state, an increase in $ N $, the number of uses of the channel, does not necessarily enhance the QFI and   it may even lead to decrease of the precision of the temperature estimation.
 We address how the optimal number of interacting qubits with which the QFI is maximized, is affected by other parameters. Our computation shows that the optimal value of $ N $, is given by

  \begin{equation}\label{NOPT}
  N^{GHZ}_{\text{opt}}=\text{R}(\dfrac{S(-2\text{e}^{-2})+2}{2\Gamma})\approx \text{R}( \dfrac{0.7968}{\Gamma}),
   \end{equation}
where $S(z)  $ denotes the principal solution for $ w $ in $ z=w\text{e}^{w} $ and R(x) rounds $ x $ to the nearest integer. If the above formula  leads to  value smaller than 3, we conclude that the best estimation occurs for N=3.

\par

 For initial GHZ state, we find that very squeezed fields needs less qubits for achieving optimal estimation of the temperature (see Fig. \ref{qfighz1}).  Nevertheless, it should be noted that
 the squeezing first leads to increase in the number of channel uses  for achieving the  optimal thermometry.
 However, more squeezing, corresponding to values of $ r $  larger than  some critical value $ r_{crit} $, reverses the process, and consequently  it decreases the number of uses of the channel.
  Moreover, 
 as shown in Fig. \ref{qfighz2}, when the relative phase $ \delta\theta $ increases from $ 0 $ to $ 2\pi $ \text{rad}, the number of uses of the channel can be reduced and  the optimal value of estimation, not affected by variation of the relative phase, is obtained with less qubits interacting with the bath. Similarly, it is found that when the S-B coupling is strengthened, the optimal $ N$ for which the QFI is maximized decreases. Therefore, if the interacting qubits are weakly coupled to the bath  the cost of quantum thermometry may increase.

On the other hand, starting from state $ |W\rangle_{N} $ such that one of the qubits is affected by the channel (n=1) and computing the spectral decomposition of the evolved density matrix, one finds that all the eigenvalues (eigenvectors) except  two vanish (are $ T $-independent), resulting in the following simple expression of the QFI  for any  choice of $N $:

\begin{equation}\label{QFIWN}
QFI_{N,1}(\rho^{W}_{\text{out}}) = \frac{2(N-1)}{N^{2}}\big(\coth(\Gamma)-1\big)  \big(\dfrac{\partial \Gamma}{\partial T}\big)^{2}~~~~~~\text{for}~~~N\geq 3. 
\end{equation}
 It is obvious  that the upper bound can never  be saturated in this situation. Hence, when one of the qubits is affected by the channel, the single-probe strategy   leads to more precise estimation than the ancilla-assisted one started from the W state:

\begin{equation}\label{compare1}
 F^{\theta_{0}=\pi/2}_{T}=QFI_{N,1}(\rho^{GHZ}_{\text{out}})\geq QFI_{N,1}(\rho^{W}_{\text{out}}).
\end{equation}

Therefore, the entangled strategy is not always more efficient than the single-qubit one for quantum thermometry.

\par
In the parallel strategy,  starting from  $ |W\rangle_{N} $ and diagonalizing the evolved density matrix, again we see
 that all the eigenvectors are $ T $-independent and all the eigenvalues  except $ N $  vanish. Hence, one can  find that the QFIs corresponding to different parallel strategies satisfy
 \begin{equation}\label{QFIWNpar}
 QFI_{N,N}(\rho^{W}_{\text{out}}) = \frac{4(N-1)}{\text{e}^{4\Gamma}+(N-2)\text{e}^{2\Gamma}-(N-1)}  \big(\dfrac{\partial \Gamma}{\partial T}\big)^{2}~~~~~~\text{for}~~~N\geq 3. 
 \end{equation}
 \begin{figure}[ht]
                    \includegraphics[width=10cm]{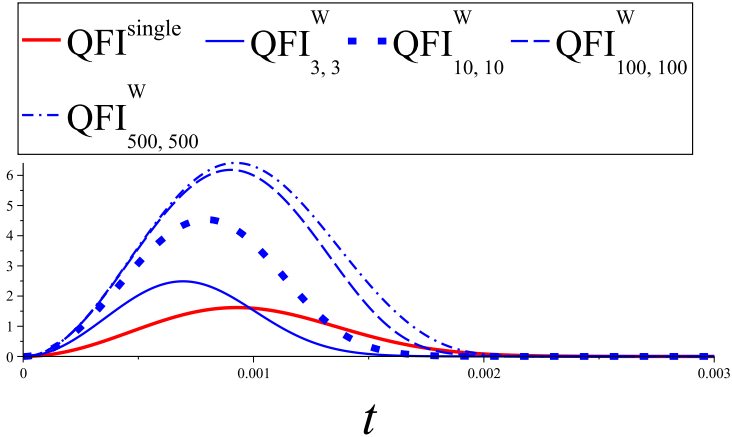}
                    \caption{\small  \small Enhanced quantum thermometry using the parallel strategy and starting from the W state. The solid red  curve
                                                illustrates the normalized single-qubit  QFI as a function of time  while the blue curves represent the normalized multi-qubit  QFIs versus time    for $ \lambda = 0.3, \delta\theta = 5, r = 0.7  $.}
                    \label{singlew}
                      \end{figure} 
 
 Plotting $ QFI_{N,N}(\rho^{W}_{\text{out}}) $ as a function of  $ N $ illustrates that the QFI is enhanced when the number of the channel uses increases. Increasing $ N $  can also raise the efficiency of the parallel strategy with initial W state  with respect to the single-qubit strategy for quantum thermometry (see Fig. \ref{singlew}). 
 
\par
For $ n>1 $, a compact expression for $ QFI_{N,n}(\rho^{W}_{\text{out}}) $ is not generally accessible except for some special cases presented here briefly:

 \begin{equation}\label{QFIW32}
 QFI_{3,2}(\rho^{W}_{\text{out}})={\frac {8({1+{\rm e}^{2\,\Gamma}})}{6\,{{\rm e}^{4\,\Gamma}}-3\,{{\rm e}^{2\,\Gamma}}-3}}
  \big(\dfrac{\partial \Gamma}{\partial T}\big)^{2},
                            \end{equation}
                            
    \begin{equation}\label{QFIW42}
      QFI_{4,2}(\rho^{W}_{\text{out}})={\frac {2({2~{\rm e}^{2\,\Gamma}+1})}{3\,{{\rm e}^{4\,\Gamma}}-2\,{{\rm e}^{2\,\Gamma}}-1}}
        \big(\dfrac{\partial \Gamma}{\partial T}\big)^{2},
           \end{equation}
    \begin{equation}\label{QFIW43}
         QFI_{4,3}(\rho^{W}_{\text{out}})={\frac {3({{\rm e}^{2\,\Gamma}}+2)}{2(\,{{\rm e}^{4\,\Gamma}}-1)}}
                 \big(\dfrac{\partial \Gamma}{\partial T}\big)^{2},
              \end{equation}
 \begin{equation}\label{QFIW62}
      QFI_{6,2}(\rho^{W}_{\text{out}})={\frac {4({4~{\rm e}^{2\,\Gamma}+1})}{15\,{{\rm e}^{4\,\Gamma}}-12\,{{\rm e}^{2\,\Gamma}}-3}}
              \big(\dfrac{\partial \Gamma}{\partial T}\big)^{2},
           \end{equation}
           \begin{equation}\label{QFIW65}
                 QFI_{6,5}(\rho^{W}_{\text{out}})={\frac {5({{\rm e}^{2\,\Gamma}+4})}{3\,{{\rm e}^{4\,\Gamma}}+3\,{{\rm e}^{2\,\Gamma}}-6}}
                               \big(\dfrac{\partial \Gamma}{\partial T}\big)^{2}.
                      \end{equation}
                      
                                  \begin{figure}[ht!]
                                              \subfigure[]{\includegraphics[width=7.1cm]{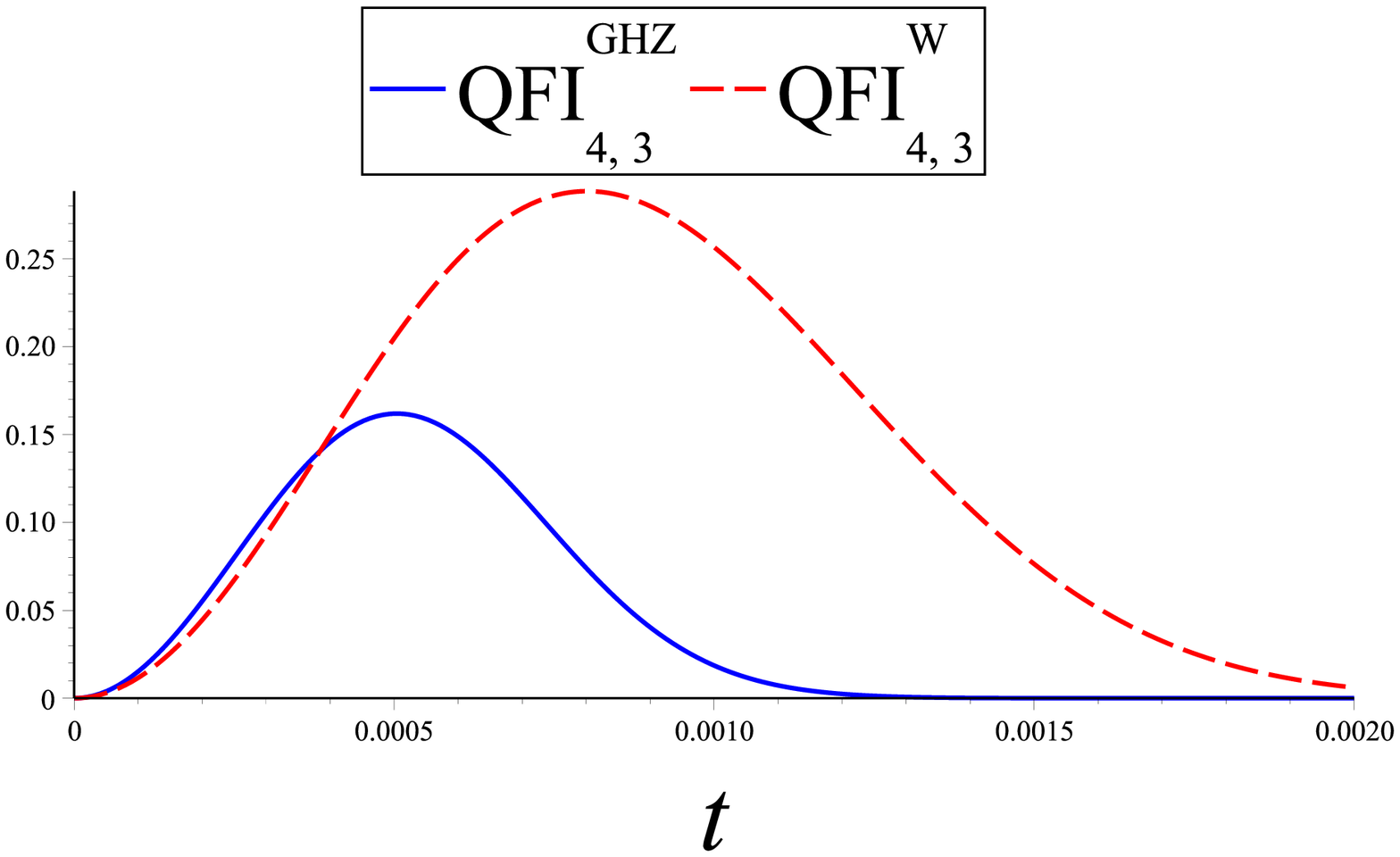}\label{QFIM43}}
                                            \subfigure[]{\includegraphics[width=7.15cm]{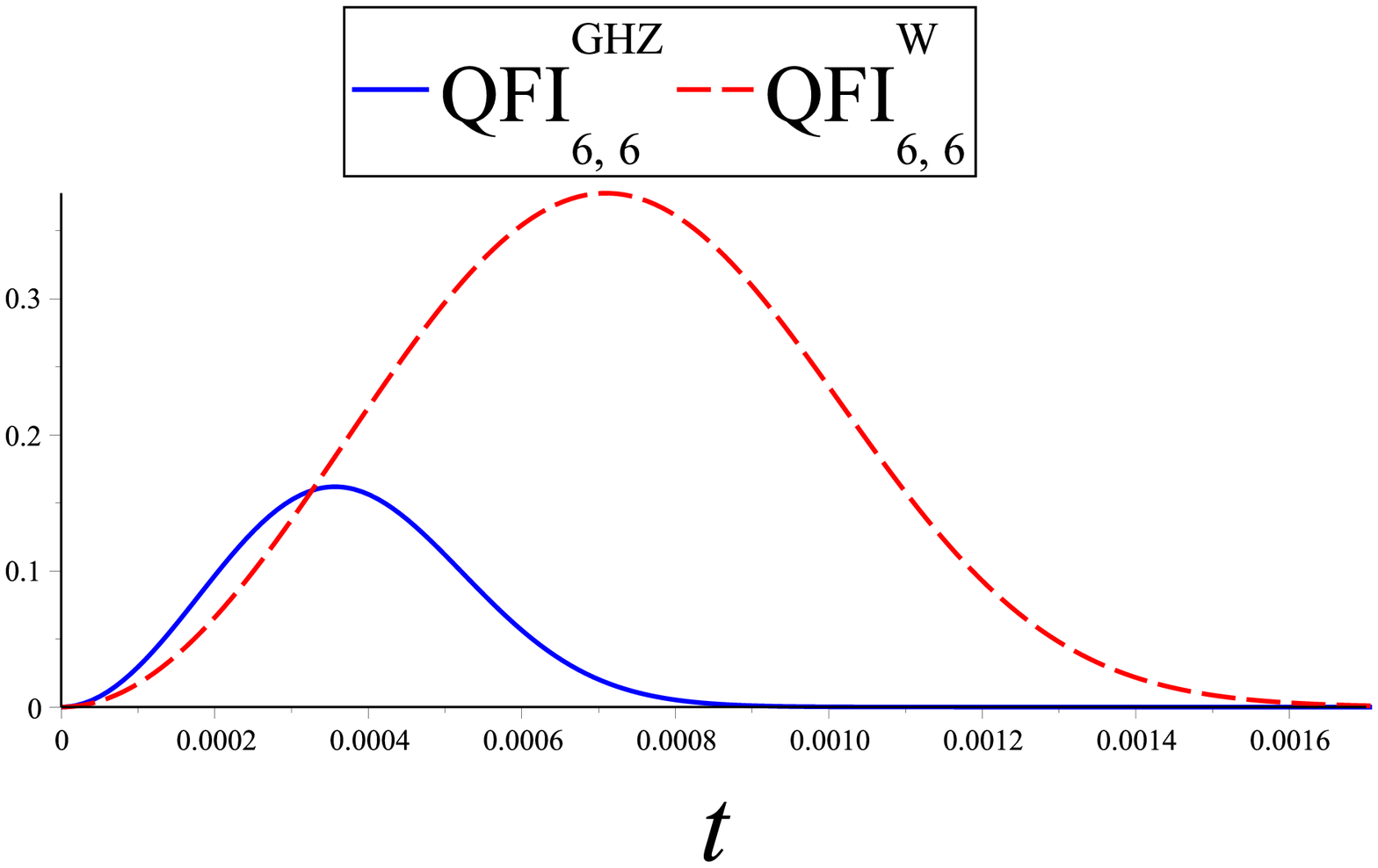}\label{QFIM66}}
                                                            \caption{\small Comparying the normalized multi-qubit QFIs  corresponding to initial GHZ and W states  for $ \lambda = 0.6, \delta\theta = 5$, and   $r =1  $.}
                                                                                                                                         \label{QFIM4366}
                                    \end{figure}
                      
    According to Eq. (\ref{compare1}), it is concluded that for $ n=1 $, starting from the GHZ state or  adopting  the single-qubit strategy, we  achieve more accurate estimation than starting from  W state. However, for  $ n>1 $,  initial preparation of the probes in W state leads to more efficient thermometry than starting with GHZ state (see Fig. \ref{QFIM4366}).        
           
      \par
         Now an important question arises: are the GHZ and W states  the optimal
         one? Are there (entangled) states which lead to more accurate estimation?
         The answer to this question cannot be analytic, since
         the analytical  diagonalization of   a multi-qubit density matrix
         is not generally possible. In order to solve this problem, we should attack it numerically and 
         first notice that the QFI is maximized for an initial
         pure state \cite{Fujiwara0423042001}. After generating  a large number
        of random initial pure states by  \textit{QUBIT4MATLAB V5.6} \cite{TothCPC}, we find that 
         the QFI, evaluated numerically using  expression (\ref{qfiasli}),   is maximized for either GHZ or W states for N = 2,3,4,5,6.
         Nevertheless,   our limited observations should
         not be understood as a certain  result, because the general answer can be presented  only  when we can check all the initial states while solving the eigenvalue problem for high-dimensional density matrices   leads to some complexity in the process of computing the QFI. Moreover,  the iterative method discussed in Ref. \cite{Macieszczak} also bypasses
         the direct maximization of the QFI,
         however, that method is only designed for those
         quantum channels that can be represented as $ \text{e}^{-iH\theta} \Lambda(\rho)\text{e}^{iH\theta} $
         (in which $ \Lambda $ denotes a noisy channel) under certain assumptions and for estimating $ \theta $.

\par
It should be noted that all of the multi-qubit QFIs   qualitatively exhibit the behaviour discussed in Figs. \ref{qfi1omlam}-\ref{qfi1delta12} for single-qubit QFI.

    \section{Summary and conclusions \label{conclusion}}
      \par To summarize, we discussed in detail the 
     quantum thermometry  by using a qubit subjected to the dephasing dynamics via interacting with 
         a squeezed thermal bath. In particular, it was investigated how we can practically design the optimal estimation for one-qubit probe. Moreover,  we illustrated that the optimum  precision  obtained  in the process of thermometry is robust against squeezing. In addition, it was shown that squeezing parameters lead to interesting and non-trivial effects on the quantum thermometry. 
 Generalizing the results for entangled probes and analysing the multi-qubit strategies  for estimating the temperature, we found that 
that in the entangled strategy with initial GHZ state, an increase in  the number of uses of the channel, does not necessarily enhance the QFI and   it may even lead to decrease of the precision of the temperature estimation. Moreover, 
         the squeezing may  decrease the thermometry costs. 
 We also addressed  how   the parallel strategy   starting from the W state becomes more efficient than  the single-qubit strategy  for quantum thermometry.

    \section{Acknowledgements}
    
     I wish to acknowledge the financial support of
     the MSRT of Iran and Jahrom University.

\pagebreak


\begin{thebibliography}{2}
\footnotesize
\bibitem{Helstron} C. W. Helstron,  \textit{Quantum detection and estimation theory} (Academic Press,
1976).

\bibitem{Israel2014} Y. Israel, S. Rosen,  and Y. Silberberg,   Phys. Rev. Lett. \textbf{112}, 103604 (2014).

\bibitem{GiovannettiPRL2006} V. Giovannetti, S. Lloyd, and L. Maccone, Phys. Rev. Lett. \textbf{96},
010401 (2006).

\bibitem{Giovannetti2004} V. Giovannetti, S. Lloyd, and L. Maccone, Science \textbf{306}, 1330
(2004).

\bibitem{Dam2007} W. van Dam, G. M. D’Ariano, A. Ekert, C. Macchiavello, and
M. Mosca, Phys. Rev. Lett. \textbf{98}, 090501 (2007).

\bibitem{M.G.A. Paris} M.G.A. Paris, Int. J. Quant. Inf. \textbf{7}, 125 (2009).
\bibitem{RanganiAOP} H. Rangani Jahromi, M. Amniat-Talab, Ann. Phys. \textbf{355}, 299 (2015).

\bibitem{RanganiAOP2} H. Rangani Jahromi and M. Amniat-Talab, Ann. Phys. \textbf{360}, 446461 (2015).
\bibitem{RanganiJMO} H. Rangani Jahromi, J. Mod. Opt. \textbf{64}, 1377 (2017).
\bibitem{RanganiOPTC} H. Rangani Jahromi, Opt. Commun. \textbf{411}, 119 (2018).
\bibitem{RanganiQIP4}  M. Jafarzadeh, H. Rangani Jahromi, and M. Amniat-Talab, Quantum Inf. Process \textbf{17},
165 (2018).
 
 \bibitem{RanganiAr1} H. Rangani Jahromi, arXiv:1807.09362.



\bibitem{Huang2018} Z. Huang, C. Macchiavello, and L. Maccone, Phys. Rev. A \textbf{97}, 032333 (2018).

\bibitem{Demkowicz2012} R. Demkowicz-Dobrza\'{n}ski, J. Kolody\'{n}ski, and M. Guta,
Nature Comm. \textbf{3}, 1063 (2012).



\bibitem{Berry2000} D. W. Berry, and H. M. Wiseman,  Phys.  Rev. Lett. \textbf{85}, 5098–5101 (2000).

\bibitem{Zwierz2010} M.Zwierz, C. A.Prez-Delgado and  P. Kok,  Phys.  Rev. Lett. \textbf{105}, 180402 (2010).
\bibitem{DemkowiczPRL2014} R. Demkowicz-Dobrza\'{n}ski, and L. Maccone, Phys.  Rev. Lett. \textbf{113}, 250801 (2014).

\bibitem{Qin LiuNCOM} G.-Q. Liu, Y.-R. Zhang, Y.-C. Chang, J.-D. Yue1, H. Fan, and X.-Y. Pan, Nature Comm. \textbf{6}, 6726 (2015).
\bibitem{Wineland1992} D. J. Wineland, W. M. Itano, J. J. Bollinger, and F. L. Moore, Phys. Rev. A \textbf{46}, R6797
(1992).

\bibitem{Huelga1997} S. F. Huelga, C. Macchiavello, T. Pellizzari, and A. K. Ekert, Phys. Rev. Lett. \textbf{79}, 3865 (1997).

\bibitem{Wasilewski2010} W. Wasilewski, K. Jensen, H. Krauter, J. J. Renema, M. V. Balabas, and E. S. Polzik, Phys. Rev. Lett. \textbf{104}, 133601 (2010).

\bibitem{Koschorreck2010} M.Koschorreck, M. Napolitano,  B.Dubost, and M. W.Mitchell, Phys. Rev. Lett. \textbf{104},
093602 (2010).

\bibitem{Mitchell2004} M. W. Mitchell, J. S. Lundeen,  and A. M.Steinberg,  Nature \textbf{429}, 161
(2004).
\bibitem{Nagata2007} T. Nagata, R.Okamoto, J. L. O'Brien, K. Sasaki, and S. Takeuchi,   Science \textbf{316}, 726
(2007).

\bibitem{LIGO} LIGO Collaboration, Nat. Phys. \textbf{7}, 962 (2011).


\bibitem{Hyllus2010} P. Hyllus, O. G\"{u}hne  and A. Smerzi,   Phys. Rev. A \textbf{82} 012337 (2010).
\bibitem{Boixo2008} S. Boixo, A. Datta, M. J. Davis, S. T. Flammia, A. Shaji, and C. M. Caves, Phys. Rev. Lett. \textbf{101} 040403 (2008).

\bibitem{Tilma2010} T. Tilma, S. Hamaji, W. J. Munro, and K. Nemoto, Phys. Rev.
A \textbf{81}, 022108 (2010).

\bibitem{Datta2012} A. Datta and A. Shaji, Mod. Phys. Lett. B \textbf{26}, 1230010 (2012).
\bibitem{Sahota2015}J. Sahota and N. Quesada, Phys. Rev. A \textbf{91}, 013808 (2015).

\bibitem{Williams2011} N. S. Williams, K. Le Hur, and A. N. Jordan, J. Phys. A: Math.
Theor. \textbf{44}, 385003 (2011).

\bibitem{Kliesch2014} M. Kliesch, C. Gogolin, M. J. Kastoryano, A. Riera, and J.
Eisert, Phys. Rev. X \textbf{4}, 031019 (2014).

\bibitem{Millen2016} J. Millen and A. Xuereb, New J. Phys. \textbf{18}, 011002 (2016).

\bibitem{Vinjanampathy2016} S. Vinjanampathy and J. Anders, Contemp. Phys. \textbf{57}, 545
(2016).
\bibitem{Mohr2005} P. J. Mohr and B. N. Taylor, Rev. Mod. Phys. \textbf{77}, 1 (2005).
\bibitem{Weng2014} W. Weng, J. D. Anstie, T. M. Stace, G. Campbell, F. N. Baynes,
and A. N. Luiten, Phys. Rev. Lett. \textbf{112}, 160801 (2014).

\bibitem{Pasquale2016} A. De Pasquale, D. Rossini, R. Fazio, and V. Giovannetti,
Nat. Commun. \textbf{7}, 12782 (2016).
\bibitem{Pasquale2017} A. De Pasquale, K. Yuasa, and V. Giovannetti, Phys. Rev. A \textbf{96},
012316 (2017).
\bibitem{Rangani2018} B. Farajollahi, M. Jafarzadeh, H. Rangani Jahromi, and M. Amniat-Talab, Quant. Inf. Proc. \textbf{17}, 119 (2018).
\bibitem{Campbell2017}S. Campbell, M. Mehboudi, G. De Chiara, and M. Paternostro,
New J. Phys. \textbf{19}, 103003 (2017).
\bibitem{Palma2017} G. De Palma, A. De Pasquale, and V. Giovannetti, Phys. Rev. A
\textbf{95}, 052115 (2017).

\bibitem{Brunelli2012} M. Brunelli, S. Olivares, M. Paternostro, and M. G. A. Paris,
Phys. Rev. A \textbf{86}, 012125 (2012).

\bibitem{Brunelli2011} M. Brunelli, S. Olivares, and M. G. A. Paris, Phys. Rev. A \textbf{84},
032105 (2011).

\bibitem{Johnson2016}  T. H. Johnson, F. Cosco, M. T. Mitchison, D. Jaksch, and S. R.
Clark, Phys. Rev. A \textbf{93}, 053619 (2016).

\bibitem{Hohmann2016} M. Hohmann, F. Kindermann, T. Lausch, D. Mayer, F. Schmidt,
and A. Widera, Phys. Rev. A \textbf{93}, 043607 (2016).

\bibitem{BOYD}  R. Boyd, \textit{Nonlinear Optics}, (Academic Press, 2008).

\bibitem{P. Neumann2013}  P. Neumann, I. Jakobi, F. Dolde, C. Burk, R. Reuter, G. Waldherr, J. Honert, T. Wolf, A. Brunner, J. H. Shim, et al., Nano Lett. \textbf{13},  2738 (2013).

\bibitem{G. Kucsko2013} G. Kucsko, P. Maurer, N. Yao, M. Kubo, H. Noh, P. Lo, H. Park, and M. Lukin, Nature \textbf{500}, 54(2013) .

\bibitem{Toyli2013} D. M. Toyli, F. Charles, D. J. Christle, V. V. Dobrovitski, and D. D. Awschalom, Proc. Natl. Acad. Sci. USA \textbf{110}, 8417 (2013).

\bibitem{Kiilerich2018} A. H. Kiilerich, A. D. Pasquale, and V. Giovannetti, Phys. Rev. A \textbf{98}, 042124 (2018).

\bibitem{Razavian2018} S. Razavian, C. Benedetti, M. Bina, Y. Akbari-Kourbolagh, Matteo G. A. Paris, arXiv:1807.11810v1.
\bibitem{Klaers2017} J. Klaers, S. Faelt, A. Imamoglu, and E. Togan, Phys. Rev. X \textbf{7}, 031044 (2017).

\bibitem{Benedetti0321142014} C. Benedetti, F. Buscemi, P. Bordone, M.G.A. Paris, Phys. Rev. A \textbf{89}, 032114 (2014).
\bibitem{Benedetti24952014} C. Benedetti, M.G.A. Paris, Phys. Lett. A \textbf{378}, 2495 (2014).
\bibitem{Rossi0103022015} M.A.C. Rossi, M.G.A. Paris, Phys. Rev. A \textbf{92}, 010302(R) (2015).
\bibitem{Javed172018} M. Javed, S. Khan, S.A. Ullah, Quantum Inf. Process. \textbf{17}, 53 (2018)
\bibitem{Kenfack11232019} L. T. Kenfack, M. Tchoffo,L. C. Fai, Phys. Lett. A \textbf{383}, 1123 (2019).









\bibitem{Landauer} R. Landauer, IBM J. Res. Dev. \textbf{5}, 183 (1961).

\bibitem{Pop2010} E. Pop, Nano Res. \textbf{3}, 147 (2010).
\bibitem{Goold2015} J. Goold, M. Paternostro, and K. Modi, Phys. Rev. Lett. \textbf{114}, 060602 (2015).
\bibitem{Manzano2018} G. Manzano, Eur. Phys. J. Spec. Top. \textbf{227}, 285 (2018).
\bibitem{Klaers2019} J. Klaers, Phys. Rev. Lett. \textbf{122}, 040602 (2019).
 \bibitem{Scully} M.  O.  Scully, and  M.  S.  Zubairy,
 \textit{Quantum  Optics}
 (Cambridge Univ. Press, 1997)
\bibitem{Rossnagel2014} J. Rossnagel, O. Abah, F. Schmidt-Kaler, K. Singer, and
E. Lutz, Phys. Rev. Lett. \textbf{112}, 030602 (2014).
\bibitem{Carnot} S. Carnot, \textit{R\'{e}flexions sur la Puissance Motrice du feu
et sur les Machines Propres a D\'{e}velopper Cette Puissance} (Bachelier, Paris, 1824).










\bibitem{Dong1} H. Dong, S.-W. Li, Z. Yi, G. S. Agarwal, and M. O. Scully,
arXiv:1608.04364.

\bibitem{Dong2} H. Dong, D.-W. Wang, and M. B. Kim, arXiv:1706.02636.

\bibitem{May2000} V. May and O. Kühn, \textit{Charge and Energy Transfer Dynamics in
Molecular Systems: A Theoretical Introduction}, 1st ed. (Wiley-VCH, Berlin, 2000).












\bibitem{Helstrom1976} C. W. Helstrom, \textit{Quantum Detection and Estimation Theory} (Academic, New York, 1976).
\bibitem{V. GiovannettiPRL} V. Giovannetti, S. Lloyd, and L. Maccone, Phys. Rev. Lett. \textbf{96}, 010401 (2006).
\bibitem{Wei Zhong} W. Zhong, Z. Sun, J. Ma, X. Wang, and F. Nori, Phys. Rev. A \textbf{87},  022337 (2013).
\bibitem{Nielson} M. A. Nielsen and I. L. Chuang, \textit{Quantum Computation and
 Quantum Information} (Cambridge University Press, Cambridge,
 2000).
 \bibitem{Eschernature} B. M. Escher, R. L. de Matos Filho, and L. Davidovich,
 Nat. Phys. \textbf{7}, 406 (2011).
 
  \bibitem{Breuer}H. Breuer and F. Petruccione, \textit{The Theory of Open Quantum
  Systems} (Oxford University Press, Oxford, 2002).
 
\bibitem{Banerjee2007} S. Banerjee and R. Ghosh, J. Phys. A: Math. Theor. \textbf{40}, 13735
 (2007).
 

 
 \bibitem{YiNingYou} Yi-Ning You and Sheng-Wen Li, Phys. Rev. A \textbf{97},  012114 (2018).
 
  \bibitem{Chapeau2015} F. Chapeau-Blondeau, Phys. Rev. A \textbf{91}, 052310 (2015).
 
 \bibitem{Geze} G. T\'{o}th and L. Apellaniz, J. Phys. A: Math. Theor. \textbf{47}, 424006 (2014).
 

 
 \bibitem{Greenberger11311990} D. M. Greenberger, M. A. Horne, A. Shimony, and A. Zeilinger, Am. J. Phys. \textbf{58}, 1131 (1990).
 
 \bibitem{Dur0203032001}  W. D\"{u}r,  Phys. Rev. A \textbf{63}, 020303(R) (2001).
 
 \bibitem{Zhao542004}    Z. Zhao, Y.-A. Chen, A.-N. Zhang, T. Yang, H. J. Briegel, and  J-W. Pan, Nature. \textbf{430}, 54
 (2004).
 \bibitem{Kempe9101999} J. Kempe,  Phys. Rev. A \textbf{60}, 910 (1999).
 
 \bibitem{Wang6372007} J. Wang, Q.  Zhang, and  C.-J.  Tang, Commun. Teor. Phys. \textbf{48}, 637 (2007).
 
 \bibitem{Liu31602011} W. Liu, Y. B.  Wang, and  Z. T. Jiang,  Opt. Commun. \textbf{284}, 3160 (2011).
 
 
 \bibitem{DHondt1732006}  E. D'Hondt, and  P. Panangaden,  Quantum Inf. Comput. \textbf{6}, 173 (2006).

 \bibitem{Mustecaplioglu} C.  B. Da\v{g},  W. Niedenzu, \"{O}.  E. M\"{u}stecaplıo\v{g}lu, and G. Kurizki, Entropy \textbf{18}, 244 (2016).
  \bibitem{TothCPC} G. T\'{o}th, Comput. Phys. Commun. \textbf{179}, 430 (2008).
 \bibitem{Fujiwara0423042001} A. Fujiwara, Phys. Rev. A \textbf{63}, 042304 (2001).
 \bibitem{Macieszczak} K. Macieszczak, arXiv:1312.1356.
 

 









   
   
   
   
   
  
\end{thebibliography}
\end{document}